\documentclass[journal]{IEEEtran}

\usepackage{cite}
\usepackage{graphicx}
\usepackage{subcaption}
\usepackage{psfrag}
\usepackage{overpic}
\usepackage[monochrome]{xcolor}

\usepackage[cmex10]{amsmath}
\usepackage{amssymb}
\usepackage{color}
\usepackage{amsthm}
\usepackage{enumitem,kantlipsum}

\usepackage[ruled,vlined,linesnumbered]{algorithm2e}
\SetCommentSty{mycommfont}

\SetKwInput{KwInput}{Input}                
\SetKwInput{KwOutput}{Output}

\def\tr{\mathrm{tr}}

\newtheorem{lemma}{Lemma}
\newtheorem{corollary}{Corollary}
\newtheorem{definition}{Definition}

\newtheorem{remark}{Remark}

\def\CN{\mathcal{N}_{\mathbb{C}}} 
\newcommand{\vect}[1]{\mathbf{#1}}
\def\tr{\mathrm{tr}}

\def\Psiv{\vect{Q}}

\def\eig{\mathrm{eig}}
\def\kron{\otimes}
\def\tr{\mathrm{tr}}

\def\Htran{\mbox{\tiny $\mathrm{H}$}}
\def\Ttran{\mbox{\tiny $\mathrm{T}$}}
\def\CN{\mathcal{N}_{\mathbb{C}}} 
\def\taupu{\tau_{\rm{p}}} 
\def\bphiu{\boldsymbol{\phi}} 

\def\Pu{\mathcal{P}} 

\begin{document}

\title{Code-domain NOMA in Massive MIMO: \\ When is it Needed?\vspace{-.2cm}}
\author{Mai T. P. Le, Luca Sanguinetti, \emph{Senior Member, IEEE}, Emil Bj{\"o}rnson, \emph{Senior Member, IEEE}, Maria-Gabriella Di Benedetto, \emph{Fellow, IEEE}\vspace{-0.6cm}

\thanks{
\newline\indent Part of this paper was presented at IEEE PIMRC 2019 \cite{MaiPIMRC19}.
Mai T.~P.~Le is with The University of Danang -- University of Science and Technology, Da Nang, Viet Nam (lpmai@dut.udn.vn). L.~Sanguinetti is with the University of Pisa, Dipartimento di Ingegneria dell'Informazione, 56122 Pisa, Italy (luca.sanguinetti@unipi.it). E.~Bj\"ornson is with the Department of Electrical Engineering (ISY), Link\"{o}ping University, 58183 Link\"{o}ping, Sweden (emil.bjornson@liu.se). M.-G. Di Benedetto is with the University of Rome ``La Sapienza'', 00184 Rome, Italy (mariagabriella.dibenedetto@uniroma1.it).}}


\maketitle

\begin{abstract}
In overloaded Massive MIMO (mMIMO) systems, wherein the number $K$ of user equipments (UEs) exceeds the number of base station antennas $M$, it has recently been shown that non-orthogonal multiple access (NOMA) can increase the sum spectral efficiency. This paper aims at identifying cases where code-domain NOMA can  improve the spectral efficiency of mMIMO in the classical regime where $K < M$. Novel spectral efficiency expressions are provided for the uplink and downlink with arbitrary spreading signatures and spatial correlation matrices. 
Particular attention is devoted to the planar arrays that are currently being deployed in pre-5G and 5G networks (in sub$-6$ GHz bands), which are characterized by limited spatial resolution. Numerical results show that mMIMO with such planar arrays can benefit from NOMA in scenarios where the UEs are spatially close to each other. 
A two-step UE grouping scheme is proposed for NOMA-aided mMIMO systems that is applicable to the spatial correlation matrices of the UEs that are currently active in each cell. Numerical results are used to investigate the performance of the algorithm under different operating conditions and types of spreading signatures (orthogonal, sparse and random sets). The analysis reveals that orthogonal signatures provide the highest average spectral efficiency.
\end{abstract}



\vspace{-0.3cm}
\section{Introduction}
Massive MIMO (mMIMO) \cite{Marzetta2010,massivemimobook} and Non-Orthogonal Multiple Access (NOMA) \cite{Dai2018Survey,islam2016power, LeIET2018} are two physical layer technologies that have received large attention in recent years. While mMIMO has already made it into the 5G standard \cite{Parkvall2017a}, the NOMA functionality remains to be standardized. Since mMIMO will likely be a mainstream feature in 5G, it is important to determine if and how NOMA can improve its performance. This is the main topic of this paper. 

\subsection{Related Work and Motivation}

Conventional multiple access schemes assign orthogonal resources to each user equipment (UE). This provides restricted/dedicated resources per UE but eliminates inter-UE interference. It is well-known that this approach is inefficient if the interference can be controlled in some other domain \cite{Dai2018Survey,islam2016power, le2018capacity}; the power and code domains are typically used for interference suppression in NOMA, while the spatial domain is used for mMIMO. While prior investigations addressed only one of these three domains, some recent works consider systems that combine NOMA and mMIMO. The vast majority of the state-of-the-art contributions in this direction investigate the performance of power-domain NOMA when combined with mMIMO (see \cite{Senel2019,Kuda2019NOMAaided,Zhang2017,deSena2019} and references therein). The gains are, however, generally limited since, to be efficient, power-domain NOMA requires UEs channels to be non-orthogonal, while a core feature of mMIMO is to make UE channels nearly orthogonal \cite{Senel2019}. 

Despite several theoretical works on code-domain NOMA with the conventional MIMO have been addressed recently \cite{Liu2019,Chi2018}, the combination of code-domain NOMA with mMIMO has received limited attention so far. The investigation in \cite{Ma2017NOMA} addresses the pilot transmission phase and analyzes two pilot structures, namely, orthogonal and superimposed deterministic pilots. It was shown that the superimposed approach achieves better performance in a high mobility environment with a large number of UEs. The uplink (UL) spectral efficiency and bit error rate performance of mMIMO with a code-domain NOMA scheme, called interleaved division multiple-access, were evaluated in \cite{Xu2017} with a low-complexity iterative data-aided channel estimation scheme and different suboptimal detection schemes, such as maximal ratio (MR) and zero-forcing (ZF) combining. In \cite{wang2019NOMA}, the authors considered the UL of an overloaded setting without any channel state information (CSI). Low density spreading signatures were applied and a blind belief propagation detector was proposed. In \cite{Liu2019MPANOMA}, the mean squared error of code-domain NOMA was considered as the performance metric of an overloaded mMIMO system.

{\color{black} The aim of this paper is to provide an analytical framework for the analysis of the combination of code-domain NOMA and classical mMIMO. Particular attention is devoted to the underloaded regime. This is motivated by the fact that a mMIMO network works properly when each BS have more antennas, $M$, than UEs, leading to an antenna-UE ratio $M/K > 1$ \cite{massivemimobook}. This makes linear UL receive combining and DL transmit precoding nearly optimal since each interfering UE contributes with relatively little interference. 

\subsection{Contributions}
{\color{black}The spectral efficiency (SE) of a classical mMIMO system grows without bound as $M \to \infty$ when the spatial correlation properties of the interfering UEs' channels are sufficiently different \cite{BjornsonHS17,Sanguinetti2019a}. Nevertheless, the SE that is achieved at any finite $M$ can potentially be improved. In particular, there might be use cases where the UEs are located close to each other, such as in public hubs like stadiums, offices in high-rise buildings, train stations, and public outdoor events, wherein the UEs' spatial channel correlation properties may be very similar and, thus, a very large number of antennas is needed to deliver acceptable performance when relying solely on the spatial processing provided by classical mMIMO.
Orthogonal time-frequency scheduling algorithms that deal with this situation are described in \cite{Huh2012a,Marzetta2016a}, but can these potentially be improved using NOMA?
The main objective of this paper is to answer a simple question: \emph{What are (if any) the potential benefits of code-domain NOMA with mMIMO in those use cases?} }

To provide some intuitions about the role that NOMA can play, Section~\ref{sec:singlecell2UEs} first considers the UL of a case study setup with a single cell, $K=2$ active UEs and perfectly known line-of-sight (LoS) propagation channels. The base station (BS) is equipped with $M=64$ antennas deployed on a uniform linear array (ULA) with half-wavelength spacing. The analysis is carried out for maximum ratio (MR) and minimum mean square error (MMSE) combining schemes for UEs that are located spatially close to each other such that the array cannot resolve the UE angles. This is known as an \textit{unfavorable propagation} scenario in the mMIMO literature \cite{massivemimobook,Marzetta2016a}. The analysis is then extended in Sections~\ref{Sec:SystemModel} and \ref{Sec:SE} to both the UL and DL of a general multicell mMIMO system with NOMA. {\color{black}Novel general SE expressions are provided (borrowing standard results from mMIMO literature) with arbitrary spreading sequences and spatial correlation matrices}, that are used to design combining and precoding schemes, and to evaluate system performance for two configurations of antenna arrays and channel models; that is, the 2D one-ring channel model for a ULA and the 3D one-ring channel model for a planar array. In Section~\ref{Sec:NumericalAnalysis}, these SEs are used to confirm the preliminary analysis of Section~\ref{sec:singlecell2UEs} for the case study setup with $M=64$ and $K=2$. To fully take advantage of NOMA in a general setup with multiple UEs, in Section~\ref{Sec:UEgrouping} we propose a per-cell UE grouping algorithm based on the $k-$means algorithm and using the chordal distance between spatial correlation matrices as a similarity score metric \cite{Ko2012}. The proposed per-cell UE grouping algorithm possibly operates in two steps and is applicable irrespective of the UE locations. If the UEs are located close to each other, the second step makes use of the Hungarian method to ensure that exactly $N$ UEs are assigned to each group such that $GN = K$, with $G$ being the total number of groups. This allows to make efficient use of spreading sequences in the network.

\subsection{Outline and notation}

The paper is organized as follows. Section~\ref{sec:singlecell2UEs} provides some intuition on why code-domain NOMA can be useful with mMIMO: a case study setup with a single-cell network, two UEs and deterministic LoS channels. Section~\ref{Sec:SystemModel} introduces a general signal model for NOMA-aided mMIMO with multicell operation, arbitrary spreading signatures and spatial correlation matrices. The achievable SEs in the UL and DL  are derived in Section~\ref{Sec:SE}, and used to select the optimal combining and precoding schemes. Numerical results are used to quantify the SEs in the case study setup and to validate the intuition provided in Section~\ref{sec:singlecell2UEs}. A UE grouping algorithm is developed in Section~\ref{Sec:UEgrouping}. The performance of NOMA-aided mMIMO is evaluated  in Section~\ref{Sec:PerformanceEvaluation} under different operating conditions. Conclusions are drawn in Section~\ref{Sec:conclusion}.

\textit{Notation}: 
We denote $[\vect{x}_i]$ and $[\vect{X}]_{i,j}$ the $i$th element of the vector $\vect{x}$ and $(i,j)$th element of the matrix $\vect{X}$, respectively. $\| \vect{x} \|^2$ denotes the $L_2$-norm of vector $\vect{x}$, i.e. $\| \vect{x} \|^2  = \sqrt{\sum_i |[\vect{x}]_i|^2}$, whereas the Frobenius norm of matrix $\vect{X}$ is denoted by $\|\vect{X}\|_F= \sqrt{\sum_{i,j} |[\vect{X}_{i,j}]|^2}$. $\vect{X}^T$, $\vect{X}^*$, $\vect{X}^{\Htran}$, $\tr{\vect{X}}$, $\mathbb{E} \{\vect{X}\}$ are the transpose, the complex conjugate, the conjugate transpose, the trace and the expectation of the matrix $\vect{X}$, respectively. The operator $\otimes$ stands for the Kronecker product. 
The circularly symmetric complex Gaussian distribution with zero mean and correlation matrix $\vect{R}$ is denoted by $\CN({0}, \vect{R})$. 


\section{A Gentle Start:\\Single-cell deployment with two UEs and LoS channels}\label{sec:singlecell2UEs}
To showcase what benefits code-domain NOMA can bring in a multi-antenna system, we consider the UL of a single-cell network where the BS is equipped with a uniform linear array of $M$ antennas with half-wavelength spacing, and receives signals simultaneously from $K=2$ single-antenna UEs. We denote by $\vect{h}_{k} \in \mathbb{C}^{M}$ for $k=1,2$ the channel between UE~$k$ and the BS. {\color{black}We further assume free-space LoS channels, leading to the following deterministic channel response \cite[Sec.~1.3.2]{massivemimobook}:
$
\vect{h}_{k} = \sqrt{\beta_k}{\bf a}_{k}(\phi_k)$
where $\beta_k$ is the large-scale fading attenuation and ${\bf a}_{k}(\phi_k) =\left[1, e^{{\mathsf{j}}\pi \sin(\phi_k)},\ldots,e^{{\mathsf{j}}\pi (M-1)\sin(\phi_k)}\right]^{\Ttran}$ is the array response vector with $\phi_k\in [0,2\pi)$ being the angle-of-arrival (AoA) from UE $k$, measured from the broadside of the BS array.} We assume that UEs use $N$-length spreading signatures for UL data transmission, where $N$ is a positive integer. We call $\vect{u}_{k} \in \mathbb{C}^{N}$ the spreading signature randomly assigned to {UE}~$k$ and assume that $\| \vect{u}_{k} \|^2  = N$. The $N\times 2$ matrix ${\bf U} = [\vect{u}_{1},\vect{u}_{2}]\in \mathbb{C}^{N\times 2}$ denotes the signature matrix. The received signal ${\bf Y}\in \mathbb{C}^{M\times N}$ for the duration of spreading signatures is
\begin{align} \label{eq:uplink-signal-model_case_study}
\vect{Y}=  s_{1}\vect{h}_{1}\vect{u}_{1}^{\Ttran} + s_{2}\vect{h}_{2}\vect{u}_{2}^{\Ttran} + \vect{N},
\end{align}
where $s_{i} \sim \CN({0}, p)$ is the data signal from UE~$i$ and $\vect{N} \in \mathbb{C}^{M \times N}$ is thermal noise with i.i.d.\ elements distributed as $\CN(0, \sigma_{\rm{ul}}^{2})$. 
 Note that, in the absence of spreading signatures, \eqref{eq:uplink-signal-model_case_study} reduces to the classical mMIMO signal model for the UL. 
 
To detect $s_{1}$ from $\vect{Y}$ in \eqref{eq:uplink-signal-model_case_study}, the BS uses the combining vector $\vect{v}_{1} \in \mathbb{C}^{MN}$, multiplied by the vectorized version of $\vect{Y}$, to obtain
\begin{align} 
\!\!\vect{v}_{1}^{\Htran} {\rm{vec}}\left({\bf Y}\right)  =  s_{1} \vect{v}_{1}^{\Htran}\vect{g}_{1}+  s_{2}\vect{v}_{1}^{\Htran}\vect{g}_{2} + \vect{v}_{1}^{\Htran}{\rm{vec}}\left({\bf N}\right),\label{eq:vector-channel-UL-processed_case_study}
\end{align}
where $\vect{g}_{k}  = {\rm{vec}}\big(\vect{h}_{k} \vect{u}_{k}^{\Ttran}\big) = \vect{u}_{k} \otimes {\vect{h}}_{k}\in \mathbb{C}^{MN}$ for $k=1,2$
is the \emph{effective channel vector}. By treating the interference as noise, the achievable SE for UE $1$ is
\begin{equation} \label{SE_NOMA}
\mathsf{SE}_1 = \frac{1}{N}\mathbb{E}_{\bf U} \left\{\log_2\left(1+\gamma_{1} \right)\right\},
\end{equation}
where $\gamma_{1} $ is the signal-to-interference-and-noise ratio (SINR)
\begin{align}\label{eq:uplink-instant-SINR_case_study}
\color{black}{\gamma_{1} 
= \frac{p |  \vect{v}_{1}^{\Htran} {\vect{g}}_{1}|^2  }{ 
p |  \vect{v}_{1}^{\Htran} {\vect{g}}_{2}|^2 +  \sigma_{\rm{ul}}^{2}{\vect{v}}_{1}^{\Htran}{{\vect{v}}_{1}} 
}}
\end{align}
and the expectation is taken with respect to the random assignment of signatures. The pre-log factor $\frac{1}{N}$ accounts for the fraction of samples used for transmitting the spreading signatures and it is smaller than $1$ as it would be the case with classical mMIMO. However, if the signatures are properly associated with the UEs, the SE can be higher. To better understand this, we now design the combiner $\vect{v}_{1}$ in \eqref{eq:vector-channel-UL-processed_case_study}, which must be selected as a function of $\{\vect{g}_{1},\vect{g}_{2}\}$, rather than $\{\vect{h}_{1},\vect{h}_{2}\}$ as  would be the case in classical mMIMO. \textcolor{black}{We assume that $\beta_1 = \beta_2=\beta$ and define the average received signal-to-noise ratio (SNR) as ${\mathsf{ SNR}}_{\rm{ul}} = \beta {p}/{\sigma_{\rm{ul}}^{2}}$.} We begin by considering the popular MR combining with perfect channel knowledge, defined as $\vect{v}_{1} = \vect{g}_{1}$, leading to 
\begin{align} 
\gamma_{1}^{\rm{MR}}
= \frac{ 1}{  |  \frac{1}{M} \vect{a}_{1}^{\Htran}(\phi_1) {\vect{a}}_{2}(\phi_2)|^2| \frac{1}{N}\vect{u}_{1}^{\Htran} {\vect{u}}_{2}|^2 +  \frac{1}{MN \mathsf{SNR}_{\rm{ul}}} 
},
\label{eq:uplink-instant-SINR_case_study_1_MR}
\end{align}
given that\footnote{$ ({\bf A}\otimes {\bf B})^{\Htran} = {\bf A}^{\Htran}\otimes {\bf B}^{\Htran}$ and $({\bf A}\otimes {\bf B})({\bf C}\otimes {\bf D}) = {\bf AC}\otimes{\bf BD}$}
{\color{black}$
\vect{g}_{1}^{\Htran} {\vect{g}}_{1} = \beta MN
$
and $$|\vect{g}_{1}^{\Htran} {\vect{g}}_{2}|^2  = \beta^2|\vect{a}_{1}^{\Htran}(\phi_1) {\vect{a}}_{2}(\phi_2)|^2|  \vect{u}_{1}^{\Htran} {\vect{u}}_{2}|^2.$$} 
We note that \cite[Sec.~1.3.2]{massivemimobook}
\begin{equation} \label{SE_mMIMO}
\frac{1}{M} \vect{a}_{1}^{\Htran}(\phi_1) {\vect{a}}_{2}(\phi_2) = 
\begin{cases}
\frac{\sin\left(M\Omega_{12}\right)}{M \sin(\Omega_{12})} & \text{if} \; \sin(\phi_1) \ne \sin(\phi_2) \\
1 & \text{if} \; \sin(\phi_1) = \sin(\phi_2)
\end{cases}
\end{equation}
with $\Omega_{12}=\pi(\sin(\phi_1) - \sin(\phi_2))/2$. 

The term $|  \frac{1}{M} \vect{a}_{1}^{\Htran}(\phi_1) {\vect{a}}_{2}(\phi_2)|^2| \frac{1}{N}\vect{u}_{1}^{\Htran} {\vect{u}}_{2}|^2$ accounts for the interference generated by UE 2 and $MN \mathsf{SNR}_{\rm{ul}}$ represents the received SNR in the absence of interference. From \eqref{SE_mMIMO}, it follows that the interference is stronger when the AoAs are similar to each other. However, if the UEs are associated to orthogonal codes/signatures (i.e., $\vect{u}_{1}^{\Htran} {\vect{u}}_{2} = 0$), the interference vanishes irrespective of the similarity of the AoAs, and the SE grows without limit as $\mathsf{SNR}_{\rm{ul}} \to \infty$. On the contrary, it saturates to $\log_2( 1 + 1/|  \frac{1}{M} \vect{a}_{1}^{\Htran}(\phi_1) {\vect{a}}_{2}(\phi_2)|^2)$ with mMIMO, due to the residual interference.

Instead of using the suboptimal MR combining, we note that $\gamma_{1}$ in \eqref{eq:uplink-instant-SINR_case_study} is a generalized Rayleigh quotient with respect to ${\bf v}_1$ and thus is maximized by the minimum mean square error (MMSE) combining vector \cite[Sec.~1.3.3]{massivemimobook}: 
\begin{align}
{\bf v}_1 = \Bigg(  \sum\limits_{i=1}^2 {\vect{g}}_{i} {\vect{g}}_{i}^{\Htran} + \frac{1}{\mathsf{SNR}_{\rm{ul}}}\vect{I}_{MN}  \Bigg)^{\!-1}\!\! {\vect{g}}_{1},
\end{align}
leading to 
\begin{align}
\gamma_{1}^{\rm{MMSE}}  &= {\vect{g}}_{1}^{\Htran}\Bigg(  {\vect{g}}_{2} {\vect{g}}_{2}^{\Htran} + \frac{1}{\mathsf{SNR}_{\rm{ul}}}\vect{I}_{MN}  \Bigg)^{\!-1}\!\!\!\!{\vect{g}}_{1}  \nonumber \\
 &\mathop=^{{(a)}} MN\mathsf{SNR}_{\rm{ul}} \Bigg(1 - \frac{|  \frac{1}{M} \vect{a}_{1}^{\Htran}(\phi_1) {\vect{a}}_{2}(\phi_2)|^2| \frac{1}{N}\vect{u}_{1}^{\Htran} {\vect{u}}_{2}|^2}{1 + \frac{1}{MN \mathsf{SNR}_{\rm{ul}}}} \Bigg)\label{eq:MMSE-max-SINR}
\end{align}
where $(a)$ follows from the matrix inversion lemma. The above SINR contains the same terms as \eqref{eq:uplink-instant-SINR_case_study_1_MR}, but has a different structure. In \eqref{eq:uplink-instant-SINR_case_study_1_MR}, $|  \frac{1}{M} \vect{a}_{1}^{\Htran}(\phi_1) {\vect{a}}_{2}(\phi_2)|^2| \frac{1}{N}\vect{u}_{1}^{\Htran} {\vect{u}}_{2}|^2$ must be interpreted as the perfomance loss due to the cancellation of the interference generated by UE 2. Similar to MR combining, this performance loss increases as the signals arrive from similar angles, but can be controlled (or even reduced to zero) by using spreading signatures.

 \begin{figure} 
        \centering \vspace{-0.7cm}
        \begin{subfigure}[b]{\linewidth} \centering 
                \includegraphics[width=1\linewidth]{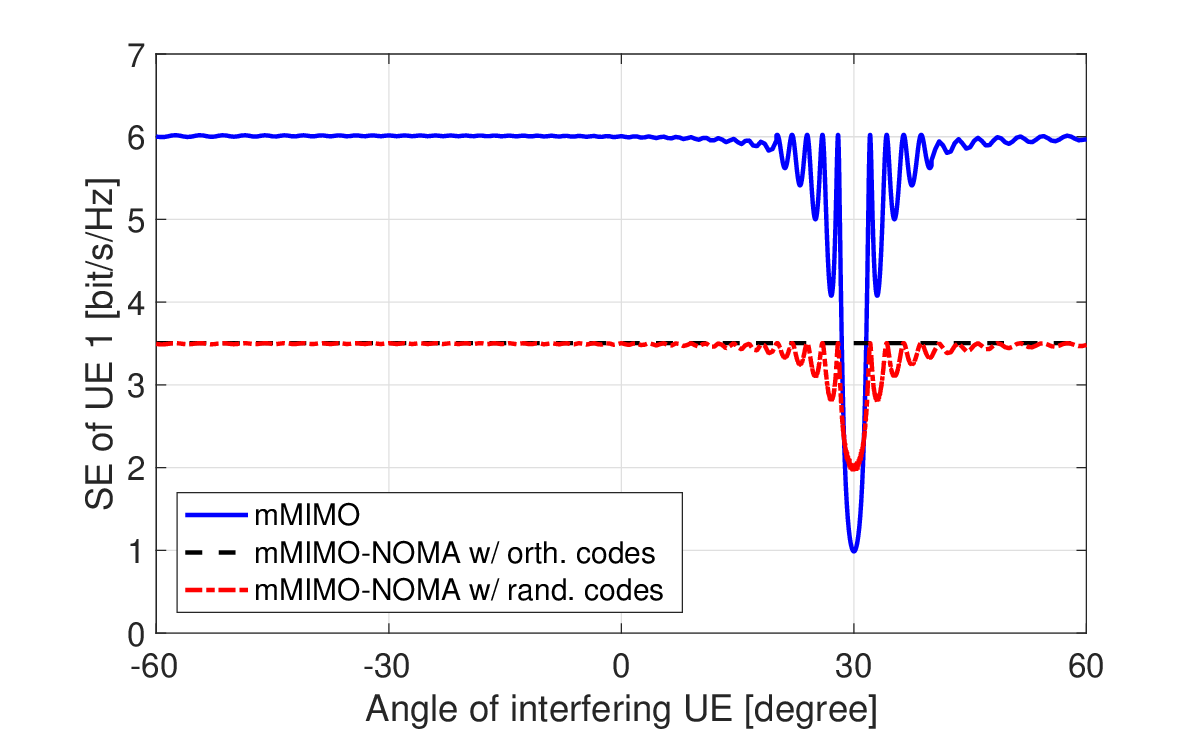} 
                \caption{MR combining} 
                \label{fig:fig_SE_toy_example_K8}
        \end{subfigure} 
        \begin{subfigure}[b]{\linewidth} \centering
                \includegraphics[width=1\linewidth]{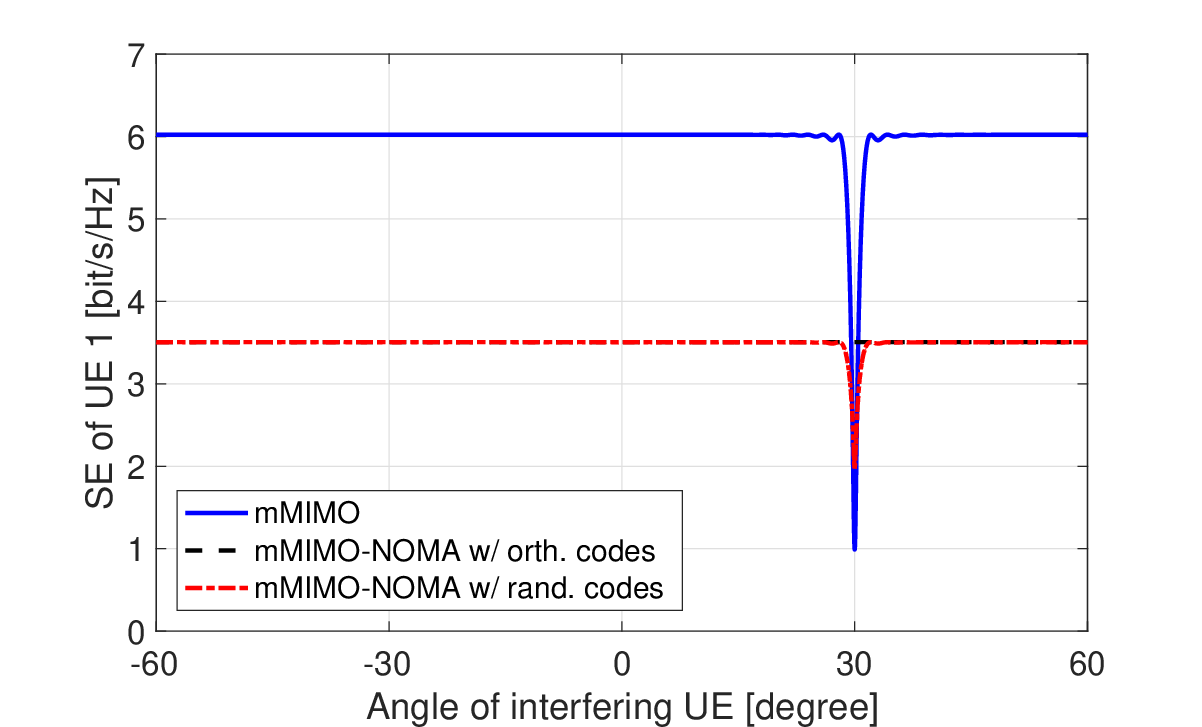} 
                \caption{MMSE combining}  
                \label{fig:fig_SE_toy_example_K32} 
        \end{subfigure}  
\vspace{-2mm}
\caption{SE of UE 1 for mMIMO-based scheme ($M=64$) with two code-domain NOMA approaches, under LoS propagation with $\phi_1 = 30^\circ$, as a function of the azimuth angle of the interfering UE. MR (Fig.\ref{fig:fig_SE_toy_example_K8}) vs. MMSE combining (Fig.\ref{fig:fig_SE_toy_example_K32}) with perfect CSI are considered.}\label{fig:fig_SE_toy_example}
\vspace{-0.1cm}
\end{figure}

To quantitatively compare the different schemes, Fig.~\ref{fig:fig_SE_toy_example} shows the SE of UE 1 when $M = 64$ and $\mathsf{SNR} = 0$ dB with MR (Fig.~\ref{fig:fig_SE_toy_example_K8}) and MMSE (Fig.~\ref{fig:fig_SE_toy_example_K32}) combining schemes. The nominal angle of UE 1 is fixed at $\phi_1 = 30^\circ$ while the angle of UE 2 varies from $-60^\circ$ to $60^\circ$. NOMA is employed with spreading signatures of length $N=2$, which are either taken from an orthogonal set or randomly picked up from an assemble of $\pm1$. Irrespective of the combining scheme and type of spreading signatures, mMIMO-NOMA outperforms mMIMO when the UEs are closely located, meaning in this case $|\phi_2 -\phi_1|\le 5^\circ$. The reason is that mMIMO is unable to spatially separate the UEs in this case. However, mMIMO achieves higher SE with both combining schemes already for $|\phi_2 -\phi_1|\ge 8^\circ$, which is a relatively small angular difference. 

{\color{black}The bottom line message of Fig.~\ref{fig:fig_SE_toy_example} is that there exist specific cases where NOMA can provide benefits if utilized with BSs equipped with many antennas $M$, even when $M\gg K$. However, several strong assumptions were made in this example; that is, single-cell operation with only 2 UEs and LoS propagation with perfect CSI. Moreover, the $64$ antennas were deployed on a large uniform linear array with half-wavelength spacing, which is unlikely to be the case in practice \cite{Sanguinetti2019a}. The question thus is: \emph{What happens in the UL and DL of practical mMIMO networks where these assumptions are not met?} }

\section{System Model}\label{Sec:SystemModel}
We consider an mMIMO network composed of $L$ cells. The BS in each cell is equipped with $M$ antennas and simultaneously serves $K$ single-antenna UEs. We assume that the BSs and UEs operate according to a TDD protocol with a data transmission phase and a pilot phase for channel estimation. We consider the standard block fading TDD protocol  \cite[Sec.~2.1]{massivemimobook} in which each coherence block consists of $\tau_{\rm c}$ channel uses, whereof $\tau_{\rm p}$ are used for UL pilots, $\tau_{\rm u}$ for UL data, and $\tau_{\rm d}$ for DL data, with $\tau_{\rm c} = \tau_{\rm p} + \tau_{\rm u} + \tau_{\rm d}$. We denote by $\vect{h}_{lk}^{j} \in \mathbb{C}^{M}$ the channel between UE~$k$ in cell~$l$ and BS~$j$. In each coherence block, an independent correlated Rayleigh fading channel realization $\vect{h}_{lk}^{j} \sim \CN \left( \vect{0}_{M}, \vect{R}_{lk}^{j}  \right)$ is drawn, where $\vect{R}_{lk}^{j} \in \mathbb{C}^{M \times M}$ is the spatial correlation matrix.
The normalized trace $\beta_{lk}^{j} =  \tr ( \vect{R}_{lk}^{j})/M$
is the average channel gain from BS~$j$ to UE~$k$ in cell~$l$. The UEs' channels are independently distributed.  {\color{black}Notice that the spatial correlation matrices $\{\vect{R}_{lk}^{j}\}$ evolve slowly in time compared to the fast variations of channel vectors $\{\vect{h}_{lk}^{j}\}$. The measurements in \cite{Viering2002a} suggest roughly two orders of magnitude slower variations. We thus assume they are available wherever needed; see \cite{BSD16A,NeumannJU17,UpadhyaJU17,CaireC17a} for practical correlation matrix estimation methods.}

\vspace{-0.2cm}
\subsection{Channel Modeling}\label{Channel_modeling}

The spatial correlation matrix $\vect{R}_{lk}^j$ describes both the array geometry and the multipath propagation environment. Models for generation of $\vect{R}_{lk}^j$ with arbitrary array geometries and environments can be found in
\cite[Sec.~7.3]{massivemimobook}.

 In this paper, we consider the following two physically motivated models:

1) \textbf{2D one-ring channel model:} This model considers a ULA with half-wavelength spacing and average path loss $\beta_{lk}^j$ \cite{Huh2012a}, \cite[Sec. 2.6]{massivemimobook}. The antennas and UEs are located in the same horizontal plane, thus the azimuth angle is sufficient to determine the directivity.
It is assumed that the scatterers are uniformly distributed in the angular interval $[\varphi_{lk}^j -\Delta, \varphi_{lk}^j + \Delta]$, where  $\varphi_{lk}^j$ is the nominal geographical angle-of-arrival (AoA) and $\Delta$ is the angular spread. This makes the $(m_1,m_2)$th element of $\vect{R}_{lk}^j$ equal to
\begin{align}\label{eq:2DChannelModel}
\left[ \vect{R}_{lk}^j \right]_{m_1,m_2} =\frac{\beta_{lk}^j}{2\Delta} \int_{-\Delta}^{\Delta}{ e^{\mathsf{j} \pi(m_1-m_2) \sin(\varphi_{lk}^j + {\varphi}) }}d{\varphi}.
\end{align}

2) \textbf{3D one-ring channel model:} This model considers a uniform planar array with the half-wavelength horizontal and vertical antenna spacing \cite[Sec. 7.3]{massivemimobook}. 
We consider a quadratic array consisting of $\sqrt M$ horizontal rows with $\sqrt M$ antennas each, which restricts $M$ to be the square of an integer.
In this case, the $(m_1,m_2)$th element of $\vect{R}_{lk}^j$ is given by
\begin{align}
&\left[ \vect{R}_{lk}^j \right]_{m_1,m_2} = \nonumber  \\
&\beta_{lk}^j \iint{ \underbrace{e^{\mathsf{j} \pi (m_1-m_2) \sin(\theta)}}_{\textrm{Vertical correlation}} \underbrace{e^{\mathsf{j} \pi (m_1-m_2) \cos(\theta)\sin(\varphi)}}_{\textrm{Horizontal correlation}} f(\varphi,\theta) d\varphi d\theta },
\label{eq:3DChannelModel}
\end{align}
 where $f(\varphi,\theta) $ is the joint probability density function of the azimuth $\varphi$ and elevation $\theta$ angles. 

 Following \cite[Sec.~7.3.2]{massivemimobook}, the 3D model is implemented by assuming that the BS height is $25$ m, the UE height is $1.5$ m, and a uniform angular distribution is used. We adopt a relative small value azimuth $\varphi = 2^{\circ}$ thorough the paper. The elevation $\theta$ of each UE is defined based on its distance to the BS of interest \cite[Sec.~7.3.2]{massivemimobook}. With fixed $\varphi = 2^{\circ}$, $\theta$ in this model ranges from about $3^{\circ}$ to about $43^{\circ}$.

{\color{black}Although the 2D model has been commonly used in the mMIMO literature (cf.~\cite{Huh2012a,Yin2013a}), the 3D model definitely better reflects the  typical pre-5G and 5G mMIMO array configurations in sub$-6$ GHz bands \cite{Bjornson2019d}. While a 64-antenna ULA can have a high angular resolution in the azimuth domain and no resolution in the elevation domain, an $8\times8$ planar array has a mediocre resolution in both domains. This might have an important impact on the spatial multiplexing capabilities, depending on where the UEs are located.}

\vspace{-0.3cm}
\subsection{Channel Estimation}

The UL pilot signature of {UE}~$k$ in cell~$j$ is denoted by the vector $\bphiu_{jk} \in \mathbb{C}^{\taupu}$ and satisfies $\| \bphiu_{jk} \|^2  = \taupu$. The elements of $\bphiu_{jk}$ are scaled by the square-root of the pilot power $\sqrt{p_{jk}}$ and transmitted over $\taupu$ channel uses, giving the received signal $\vect{Y}_j^{\rm{p}} \in \mathbb{C}^{M \times \taupu}$ at {BS}~$j$:\begin{align} \label{eq:uplink-pilot-model}
\vect{Y}_j^{\rm{p}} = \underbrace{ \sum_{i=1}^{K} \sqrt{p_{ji}} \vect{h}_{ji}^{j} \bphiu_{ji}^{\Ttran}  }_{\textrm{Desired pilots}} + \underbrace{\sum_{l=1,l \neq j}^{L} \sum_{i=1}^{K} \sqrt{p_{li}}  \vect{h}_{li}^{j} \bphiu_{li}^{\Ttran}  }_{\textrm{Inter-cell pilots}} + \underbrace{ \vphantom{\sum_{l=1,l \neq j}^{L} } \vect{N}_{j}^{p}}_{\textrm{Noise}},
\end{align}
where $\vect{N}_{j}^{p} \in \mathbb{C}^{M \times \taupu}$ is noise with i.i.d.\ elements distributed as $\CN(0,\sigma_{\rm{ul}}^{2})$. Note that we are not assuming mutually orthogonal pilot signatures, but arbitrary spreading signatures. {Hence, the MMSE estimator of $\vect{h}_{jk}^{j}$ takes a more complicated form than in prior works (e.g., \cite[Sec.~3.2]{massivemimobook}), and is given by (see Appendix A)}
\begin{equation} \label{eq:MMSEestimator_h_jli}
\widehat{\vect{h}}_{li}^{j}  = \sqrt{p_{li}}\left({\bphiu}_{li}^{\Htran} \otimes {\vect{R}}_{li}^j\right) \big(\Psiv_{li}^{j}\big)^{-1}  {\rm{vec}}\left({\bf Y}_{j}^{p}\right),
\end{equation}
with $\Psiv_{li}^{j} =  \sum_{l^\prime=1}^{L}\sum_{i^\prime=1}^{K}  p_{l^\prime i^\prime}\left({\bphiu}_{l^\prime i^\prime}{\bphiu}_{l^\prime i^\prime}^{\Htran} \right)\otimes {\vect{R}}_{l^\prime i^\prime}^j+ \sigma_{\rm{ul}}^{2}{\bf I}_{M\tau_{\rm p}}$.
The estimation error $\tilde{\vect{h}}_{li}^{j} = \vect{h}_{li}^{j} - \widehat{\vect{h}}_{li}^{j}$ is independent of $\widehat{\vect{h}}_{li}^j$ and has correlation matrix $
\vect{C}_{li}^{j} = \mathbb{E} \{ \tilde{\vect{h}}_{li}^{j} ( \tilde{\vect{h}}_{li}^{j} )^{\Htran} \}  = \vect{R}_{li}^{j} - \vect{\Phi}_{li}^j
$ with 
\begin{equation}\label{eq:Phi-definition1}
\vect{\Phi}_{li}^j ={p_{li}}\left({\bphiu}_{li}^{\Htran} \otimes {\vect{R}}_{li}^j\right)
{(\Psiv_{li}^{j})}^{-1}  \left({\bphiu}_{li} \otimes {\vect{R}}_{li}^j\right).
\end{equation} 
Note that the MMSE estimate in \eqref{eq:MMSEestimator_h_jli} holds for any choice of pilot signatures $\{\bphiu_{li}\}$, that can be arbitrarily taken from orthogonal, non-orthogonal, random, or sparse sets. In classical mMIMO, orthogonal pilot signatures are usually employed, leading to the simplified MMSE estimation expression $\widehat{\vect{h}}_{li}^{j}  = \sqrt{p_{li}}\vect{R}_{li}^{j} \big(\Psiv_{li}^{j}\big)^{-1}  \left(\vect{Y}_j^{\rm{p}}\bphiu_{li}\right)$ \cite[Sec.~3.2]{massivemimobook},
where $\Psiv_{li}^{j} =  \sum_{(l',i') \in \Pu_{li} }{p_{l'i'}}\tau_{\rm p}\vect{R}_{l'i'}^{j}  +  \sigma_{\rm{ul}}^2  \vect{I}_{M}$
and $\Pu_{li}$ collects the indices of UEs that utilize the same pilot as UE $i$ in cell $l$.

\vspace{-0.3cm}
\subsection{UL and DL data transmissions}

While classical mMIMO only uses spreading signatures for UL pilot transmission, mMIMO with NOMA utilizes $N$-length spreading signatures also for UL data transmission, $N$ being a positive integer.
We denote by $\vect{u}_{jk} \in \mathbb{C}^{N}$ the spreading signature assigned to {UE}~$k$ in cell~$j$ and assume that $\| \vect{u}_{jk} \|^2  = N$. {As for pilot transmission, the spreading signatures $\{\vect{u}_{jk}\}$ are also selected from an arbitrary set and different options will be compared below.} The received signal ${\bf Y}_{j}\in \mathbb{C}^{M\times N}$ at BS $j$ for the duration of a spreading signature is given by
\begin{align} \label{eq:uplink-signal-model}
\vect{Y}_j = \underbrace{ \sum_{i=1}^{K} s_{ji}\vect{h}_{ji}^{j}\vect{u}_{ji}^{\Ttran}}_{\textrm{Intra-cell signals}} + \underbrace{\sum_{l=1,l \neq j}^{L} \sum_{i=1}^{K}  s_{li}\vect{h}_{li}^{j}\vect{u}_{li}^{\Ttran}}_{\textrm{Inter-cell interference}} + \underbrace{\vphantom{\sum_{i=1,i\ne k}^{K} } \vect{N}_{j}}_{\textrm{Noise}},
\end{align}
where $s_{li} \sim \CN({0}, p_{li})$ is the data signal from UE~$i$ in cell~$l$ with $p_{{li}}$ being the transmit power and $\vect{N}_{j} \in \mathbb{C}^{M \times N}$ is thermal noise with i.i.d.\ elements distributed as $\CN(0, \sigma_{\rm{ul}}^{2})$. 

In the DL, the transmitted signal $\vect{X}_j \in \mathbb{C}^{M\times N}$ is given by $\vect{X}_j =  \sum_{i=1}^{K} \varsigma_{ji} \vect{W}_{ji}$
where $\varsigma_{jk} \sim \CN({0}, \rho_{jk})$ is the data signal intended for UE~$k$ in cell~$j$ and $\vect{W}_{ji}\in \mathbb{C}^{M\times N}$ is the corresponding precoding matrix that determines the spatial directivity of the signal. The received signal $\vect{y}_{jk}\in \mathbb{C}^{N\times 1}$ at UE $k$ in cell $j$, during the transmission of a spreading signature, is
\begin{align} \label{eq:downlink-signal-model_UE}
\vect{y}_{jk}^{\Htran} = \sum_{i=1}^{K} \varsigma_{ji} (\vect{h}_{jk}^j)^{\Htran}\vect{W}_{ji} + \sum_{l=1,l\ne j}^{L}\sum_{i=1}^{K} \varsigma_{li} (\vect{h}_{jk}^j)^{\Htran}\vect{W}_{li} + \vect{n}_{jk}^{\Htran},
\end{align}
where $\vect{n}_{jk} \in \mathbb{C}^{N \times 1}$ is thermal noise with i.i.d.\ elements distributed as $\CN(0, \sigma_{\rm{dl}}^{2})$. No a priori assumption is made on the precoding matrices $\{\vect{W}_{ji}\}$. In Section~\ref{DL_Spectral_Efficiency}, they will be designed based on channel estimates as well as spreading signatures used at the UEs for detection. 


\section{Spectral Efficiency}\label{Sec:SE}

In this section, we will compute the SEs that are achieved in the UL and DL when arbitrary spreading signatures are used and we will design the combining/precoding vectors.

\subsection{UL Spectral Efficiency}
To detect the data signal $s_{jk}$ from $\vect{Y}_j$ in \eqref{eq:uplink-signal-model}, BS~$j$ selects the combining vector $\vect{v}_{jk} \in \mathbb{C}^{MN}$, which is multiplied with the vectorized version of $\vect{Y}_j$ to obtain
\begin{align}
\!\!\vect{v}_{jk}^{\Htran} {\rm{vec}}\left({\bf Y}_{j}\right) 
&=  s_{jk} \vect{v}_{jk}^{\Htran}\vect{g}_{jk}^{j} + \underbrace{ \sum_{i=1,i\ne k}^{K} s_{ji}\vect{v}_{jk}^{\Htran}\vect{g}_{ji}^{j}}_{\textrm{Intra-cell interference}} \nonumber \\
 &+ \underbrace{\sum_{l=1,l \neq j}^{L} \sum_{i=1}^{K}s_{li} \vect{v}_{jk}^{\Htran}\vect{g}_{li}^{j}}_{\textrm{Inter-cell interference}} + \underbrace{\vphantom{\sum_{i=1,i\ne k}^{K} } \vect{v}_{jk}^{\Htran}{\rm{vec}}\left({\bf N}_{j}\right)}_{\textrm{Noise}},\label{eq:vector-channel-UL-processed}
\end{align}
where $\vect{g}_{li}^{j}  = {\rm{vec}}\big(\vect{h}_{li}^{j} \vect{u}_{li}^{\Htran}\big) \in \mathbb{C}^{MN}$ or, equivalently, 
\begin{align}\label{effective_channel_vec}
\vect{g}_{li}^{j} =\vect{u}_{li} \otimes \vect{h}_{li}^{j}=\left(\vect{u}_{li} \otimes {\bf I}_{M}\right)\vect{h}_{li}^{j},
\end{align}
is the effective channel vector with correlation matrix $\mathbb{E} \{ \vect{g}_{li}^{j} (\vect{g}_{li}^{j} )^{\Htran} 
\} = \left(\vect{u}_{li} \otimes {\bf I}_{M}\right)\vect{R}_{li}^{j} \left(\vect{u}_{li}^{\Htran} \otimes {\bf I}_{M}\right) = \left(\vect{u}_{li}\vect{u}_{li}^{\Htran}\right)\otimes \vect{R}_{li}^{j}$.
The MMSE estimate of $\vect{g}_{li}^{j}$ is obtained as $\widehat{\vect{g}}_{li}^{j} = \vect{u}_{li} \otimes \widehat{\vect{h}}_{li}^{j}=\left(\vect{u}_{li} \otimes {\bf I}_{M}\right)\widehat{\vect{h}}_{li}^{j}$.

Note that \eqref{eq:vector-channel-UL-processed} is mathematically equivalent to the signal model of a classical mMIMO system where the effective channel vectors are distributed as $\vect{g}_{lk}^{j} \sim \CN \left( \vect{0}_{M}, \left(\vect{u}_{lk}\vect{u}_{lk}^{\Htran}\right)\otimes \vect{R}_{li}^{j}  \right)
$ and the effective channel estimates are distributed as $\widehat{\vect{g}}_{lk}^{j} \sim \CN \left( \vect{0}_{M}, \left(\vect{u}_{lk}\vect{u}_{lk}^{\Htran}\right)\otimes\vect{\Phi}_{lk}^j  \right)$ with $\vect{\Phi}_{lk}^j$ given by \eqref{eq:Phi-definition1}. The key difference is the presence of the spreading signatures (used for UL pilot and data transmissions) in the distributions. The ergodic capacity in UL can thus be evaluated by using the well-established lower bounds developed in the mMIMO literature \cite{massivemimobook}. 
\begin{lemma}\label{theorem:uplink-SE}
If the MMSE estimator is used, an UL SE of UE $k$ in cell $j$ is
\begin{equation} \label{eq:uplink-rate-expression-general}
\begin{split}
\mathsf{SE}_{jk}^{\rm {ul}} = \frac{1}{N}\frac{\tau_{\rm u}}{\tau_{\rm c}} \mathbb{E} \left\{ \log_2  \left( 1 + \gamma_{jk}^{\rm {ul}}  \right) \right\} \quad \textrm{[bit/s/Hz] },
\end{split}
\end{equation}
where the effective instantaneous SINR $\gamma_{jk}^{\rm {ul}}$ is given in 
\begin{align}
\gamma_{jk}^{\rm {ul}} 
= \frac{ p_{jk}|  \vect{v}_{jk}^{\Htran} \widehat{\vect{g}}_{jk}^j |^2  }{ 
 \vect{v}_{jk}^{\Htran}  \left(   \sum\limits_{\substack{l=1 \\ l\ne j}}^L\sum\limits_{i=1}^K p_{li}\widehat{\vect{g}}_{li}^j {(\widehat{\vect{g}}_{li}^j)}^{\Htran}+\sum\limits_{\substack{i=1 \\ i\ne k}}^K p_{ji}\widehat{\vect{g}}_{ji}^j {(\widehat{\vect{g}}_{ji}^j)}^{\Htran} +   {\vect{Z}}_j \right) \vect{v}_{jk}  
}   \label{eq:uplink-instant-SINR}
\end{align}
with ${\vect{Z}}_j = \sum_{l=1}^{L} \sum_{i=1}^{K} p_{li}\left(\vect{u}_{li}\vect{u}_{li}^{\Htran}\right) \otimes\vect{C}_{li}^{j} +\sigma_{\rm{ul}}^{2} {\bf I}_{MN}$. {\color{blue}The expectation is taken with respect to the realizations of the effective channels, i.e., ${\vect{g}}_{li}^j = \vect{u}_{li}\kron {\vect{h}}_{li}^j$.}\footnote{{\color{blue}This is different from \eqref{SE_NOMA} in the case study of Section~\ref{sec:singlecell2UEs}, where the expectation is only taken with respect to the random assignment of signatures since the channel responses are deterministic under LoS propagation.}}
\end{lemma}
\begin{IEEEproof}
\textcolor{red}{The proof follows the same steps as that of \cite[Th. 4.1]{massivemimobook} for the signal model in~\eqref{eq:vector-channel-UL-processed} and is hence omitted.}
\end{IEEEproof}
Unlike the case study example of Section~\ref{sec:singlecell2UEs} where perfect CSI was assumed, the pre-log factor $\frac{1}{N}\frac{\tau_{\rm u}}{\tau_{\rm c}}$ in \eqref{eq:uplink-rate-expression-general} accounts for the fraction of samples used for transmitting pilot and data signatures. Whenever $N>1$, it is still smaller than $\frac{\tau_{\rm u}}{\tau_{\rm c}}$, which would be the case with classical mMIMO. 

The SE expression in \eqref{eq:uplink-rate-expression-general} holds for any combining vector and choice of spreading signatures in the data transmission. MR combining with $\vect{v}_{j k} = \widehat{\vect{g}}_{jk}^j$
is a possible choice. Similar to \eqref{eq:uplink-instant-SINR_case_study}, the expression in \eqref{eq:uplink-instant-SINR} has also the form of a generalized Rayleigh quotient. Thus, the vector that maximizes the SINR can be obtained as stated by the following lemma.

\begin{lemma} The {SINR} in \eqref{eq:uplink-instant-SINR} is maximized by
\begin{align} \label{eq:MMSE-combining}
\vect{v}_{j k} =p_{jk}\Bigg(  \sum\limits_{l=1}^L \sum\limits_{i=1}^Kp_{li}\widehat{\vect{g}}_{li}^j {(\widehat{\vect{g}}_{li}^j)}^{\Htran} + \vect{Z}_j  \Bigg)^{\!-1}    \widehat{\vect{g}}_{jk}^j,
\end{align}
leading to $\gamma_{jk}^{\rm {ul}}  = p_{jk}{(\widehat{\vect{g}}_{jk}^j)}^{\Htran}\!\!\left(  \!\sum\limits_{(l,i)\neq (j,k)}\!\!\!\!p_{li}\widehat{\vect{g}}_{li}^j {(\widehat{\vect{g}}_{li}^j)}^{\Htran} + \vect{Z}_j  \right)^{\!-1}   \!\! \!\!\widehat{\vect{g}}_{jk}^j$.
\end{lemma}
\begin{IEEEproof}
\textcolor{red}{This result follows from \cite[Lemma B.10]{massivemimobook} by replacing the channel estimates ${\widehat {\vect{h}}}_{li}^j$ with those of the effective channels, i.e., ${\widehat{\vect{g}}}_{li}^j = \vect{u}_{li}\kron {\widehat {\vect{h}}}_{li}^j$.}
\end{IEEEproof}

{\color{blue}The combining vector $\vect{v}_{j k}$ in \eqref{eq:MMSE-combining} is a function of the effective MMSE estimates $\{\widehat{\vect{g}}_{li}^j = \vect{u}_{li}\kron {\widehat{\vect{h}}}_{li}^j\}$, rather than $\{\widehat{\vect{h}}_{li}^j\}$ as would be the case in classical mMIMO. Different spreading signatures have an impact on its structure and on the corresponding SE.}
We call it \emph{NOMA MMSE (N-MMSE) combining} since it also minimizes the mean-squared error (MSE) ${\rm{MSE}}_k = \mathbb{E} \{ | s_{jk} - \vect{v}_{jk}^{\Htran} {\rm{vec}}\left({\bf Y}_{j}\right)   |^2  \big| \{ \widehat{\vect{g}}_{li}^j \}  \}$, that represents the conditional MSE between the data signal $s_{jk}$ and the received signal $\vect{v}_{jk}^{\Htran} {\rm{vec}}\left({\bf Y}_{j}\right)$, after receive combining. 

So far, we have not taken into account the structure of spreading signatures $\{\vect{u}_{jk}\}$, thus the SE expressions hold for any set of signatures. 
 We will now consider the special case when the signatures are selected from a set of mutually orthogonal vectors. In this case, the estimate of $s_{jk}$ at BS $j$ is obtained by first correlating ${\bf Y}_{j}$ with the spreading signature $\vect{u}_{jk}$ and then by multiplying the processed data signal\footnote{The processed signal ${\bf Y}_{j}\vect{u}_{jk}$ is a sufficient statistic for estimating $s_{jk}$ when the signatures are selected from a set of mutually orthogonal vectors, since there is no loss in useful information as compared to using ${\bf Y}_{j}$; see e.g.  \cite[App.~C.2.1]{massivemimobook}.} 
${\bf Y}_{j}\vect{u}_{jk}\in \mathbb{C}^{M}$ by the combining vector $\bar{\vect{v}}_{jk} \in \mathbb{C}^{M}$. 
We let $\mathcal{C}_{jk}$ denote the set of the indices of all UEs that utilize the same spreading signature as UE $k$ in cell $j$. It can be easily shown that the SINR is maximized by
\begin{align} \label{eq:MMSE-combining-orthogonal}
\bar{\vect{v}}_{j k} =p_{jk}\Bigg(  \sum\limits_{(l,i)\in\mathcal{C}_{jk}}p_{li}\widehat{\vect{h}}_{li}^j {(\widehat{\vect{h}}_{li}^j)}^{\Htran} + \bar{\vect{Z}}_{jk}  \Bigg)^{\!-1}    \widehat{\vect{h}}_{jk}^j,
\end{align}
with $
\bar{\vect{Z}}_{jk} \!= \!\!\!\!\sum\limits_{(l,i)\in\mathcal{C}_{jk}} p_{li}  \vect{C}_{li}^{j} +\frac{\sigma_{\rm{ul}}^{2}}{N} {\bf I}_{M}$ and maximum SINR $$\gamma_{jk}^{\rm {ul}}  = p_{jk}{(\widehat{\vect{h}}_{jk}^j)}^{\Htran}\!\!\left(  \sum\limits_{(l,i)\in\mathcal{C}_{jk}}\!\!\!\!p_{li}  \widehat{\vect{h}}_{li}^j {(\widehat{\vect{h}}_{li}^j)}^{\Htran} + \bar{\vect{Z}}_{jk} \right)^{\!-1}   \!\! \!\!\widehat{\vect{h}}_{jk}^j.$$

\begin{figure*}
\begin{align} \notag
\!\!\!{\gamma}_{jk}^{\mathrm{dl}} =  \frac{\rho_{jk} p_{jk} \tau_p \tr \! \left( \vect{R}_{jk}^{j} {(\Psiv_{jk}^{j})}^{-1}\vect{R}_{jk}^{j} \right)  }{\!\!\! 
\underbrace{ \vphantom{\sum\limits_{(l,i) \in \Pu_{jk}  \setminus  \{j\} }}  \sum\limits_{(l,i)\in \mathcal{C}_{jk}}\rho_{li}  \frac{ \tr \!\left( \vect{R}_{jk}^{l}  \vect{R}_{li}^{l} {(\Psiv_{li}^{l})}^{-1}   \vect{R}_{li}^{l} \right) \! }{\tr \!\left( \vect{R}_{li}^{l} {(\Psiv_{li}^{l})}^{-1}  \vect{R}_{li}^{l} \right) }}_{\textrm{Non-coherent interference}}
+ \underbrace{\!\!\!\!\sum\limits_{(l,i) \in \{\Pu_{jk}\cap\mathcal{C}_{jk}  \setminus  (j,k) \}} \!\!\!\!\!\!\frac{  \rho_{li}p_{jk}\tau_p\left| \tr  \left( \vect{R}_{jk}^{l} {(\Psiv_{li}^{l})}^{-1}  \vect{R}_{li}^{l} \right)  \right|^2  }{\tr  \left( \vect{R}_{li}^{l} {(\Psiv_{li}^{l})}^{-1}  \vect{R}_{li}^{l} \right) }}_{\textrm{Coherent interference}}+ \frac{\sigma_{\rm{dl}}^2}{N}  }.\!\!\!
\label{eq:downlink-rate-expression-forgetbound-MR-orthogonal} \tag{30}
\end{align}
\hrule
\end{figure*}

\subsection{DL Spectral Efficiency}\label{DL_Spectral_Efficiency}
We assume that, to detect the data signal $\varsigma_{ji}$ from $\vect{y}_{jk}$ in \eqref{eq:downlink-signal-model_UE}, UE $k$ in cell $j$ correlates $\vect{y}_{jk}$ with its associated spreading signature ${\bf {u}}_{jk}$ to obtain
\begin{align} \label{eq_received_signal_DL0}
z_{jk} &= \vect{y}_{jk}^{\Htran}{\bf {u}}_{jk} = (\vect{h}_{jk}^j)^{\Htran}\vect{W}_{jk} {\bf {u}}_{jk} \varsigma_{jk} +\sum_{i=1,i\ne k}^{K} (\vect{h}_{jk}^j)^{\Htran}\vect{W}_{ji} {\bf {u}}_{jk} \varsigma_{ji} \nonumber \\
&+ \sum_{l=1,l\ne j}^{L}\sum_{i=1}^{K} (\vect{h}_{jk}^j)^{\Htran}\vect{W}_{li} {\bf {u}}_{jk} \varsigma_{li} + \vect{n}_{jk}^{\Htran} {\bf {u}}_{jk}.
\end{align}
{\color{black}Notice that the UE does not know the precoded channels $(\vect{h}_{jk}^j)^{\Htran}\vect{W}_{li}$ since no pilots are transmitted in the DL. To mitigate the interference of the other UEs, it can only use its assigned spreading signature ${\bf {u}}_{jk}$.}
We denote the vectorized version of $\vect{W}_{li}$ as $\vect{w}_{li}={\rm{vec}}(\vect{W}_{li})\in \mathbb{C}^{MN}$  and observe that
\begin{align}
\!\!\!(\vect{h}_{jk}^j)^{\Htran}\vect{W}_{li} {\bf {u}}_{jk} = \left(\vect{u}_{jk} \otimes \vect{h}_{jk}^{j}\right)^{\Htran} \!\!{\rm{vec}}(\vect{W}_{li}) = (\vect{g}_{jk}^{j})^{\Htran}\vect{w}_{li}.
\end{align}
Hence, $z_{jk}$ reduces to 
\begin{align} 
\!\!\!\!\!\!z_{jk} & = (\vect{g}_{jk}^j)^{\Htran}\vect{w}_{jk} \varsigma_{jk} +\sum_{i=1,i\ne k}^{K} (\vect{g}_{jk}^j)^{\Htran}\vect{w}_{ji} \varsigma_{ji} \nonumber \\
&+ \sum_{l=1,l\ne j}^{L}\sum_{i=1}^{K} (\vect{g}_{jk}^j)^{\Htran}\vect{w}_{li}  \varsigma_{li} + \vect{n}_{jk}^{\Htran} {\bf {u}}_{jk}.\label{eq_received_signal_DL}
\end{align}
{\color{black}As in the UL, \eqref{eq_received_signal_DL} is mathematically equivalent to the signal model of classical mMIMO.} \textcolor{red}{Characterizing the capacity is harder in the DL than in the UL since it is unclear how the UE should best estimate the effective precoded channel $(\vect{g}_{jk}^j)^{\Htran}\vect{w}_{jk}$ needed for decoding. However, an achievable SE can be computed using the so-called hardening capacity bound, which has received great attention in the mMIMO literature \cite[Sec.~4.3]{massivemimobook} and will be adopted here as well.\footnote{\color{red}The hardening bound is a standard information theoretic tool for the analysis of the capacity in the DL where channel state information is not available at the UE side. The practical and theoretical implications of these bounds can be found in mMIMO textbooks (e.g., \cite[Sec. 4.3]{massivemimobook}).}}

\begin{lemma}\label{theorem:DL-SE}
The DL ergodic channel capacity of UE $k$ in cell $j$ in mMIMO-NOMA is lower bounded by
\begin{equation} \label{eq:downlink-rate-expression-general}
\begin{split}
\mathsf{SE}_{jk}^{\rm {dl}} = \frac{1}{N}\frac{\tau_{\rm d}}{\tau_{\rm c}}  \log_2  \left( 1 + \gamma_{jk}^{\rm {dl}}  \right)  \quad \textrm{[bit/s/Hz], }
\end{split}
\end{equation}
where the effective SINR $\gamma_{jk}^{\rm {dl}}$ is given as
\begin{align}
\gamma_{jk}^{\rm {dl}} 
& =  \frac{ \rho_{jk}| \mathbb{E} \{\vect{w}_{jk}^{\Htran} \vect{g}_{jk}^j \} |^2  }
{ 
\!\sum\limits_{l=1}^L\sum\limits_{i=1}^K \rho_{li} \mathbb{E} \{ | \vect{w}_{li}^{\Htran} {\vect{g}}_{jk}^l |^2 \} - \rho_{jk}| {\mathbb{E}} \{ \vect{w}_{jk}^{\Htran} {\vect{g}}_{jk}^j \} |^2
+  \sigma_{\rm{dl}}^{2}
}. \label{eq:downlink-instant-SINR}
\end{align}
The expectations are with respect to the realizations of the effective channels ${\vect{g}}_{li}^j = \vect{u}_{li}\kron {\vect{h}}_{li}^j$ $\forall j,l,i$.
\end{lemma}

\begin{IEEEproof}
\textcolor{red}{The proof follows the same steps as that of \cite[Th. 4.6]{massivemimobook} for the signal model in~\eqref{eq_received_signal_DL} and is hence omitted.}
\end{IEEEproof}

As in the UL, the DL SE in \eqref{eq:downlink-rate-expression-general} holds for any choice of precoding vectors and spreading signatures. Moreover, the pre-log factor is reduced by a factor $N$ compared to what it would be in classical mMIMO (i.e., ${\tau_{\rm d}}/{\tau_{\rm c}}$). Unlike the UL, optimal precoding design is a challenge since \eqref{eq:downlink-rate-expression-general} depends on the precoding vectors $\{{{\bf w}_{li}}\}$ of all UEs. A common heuristic approach relies on the UL-DL duality \cite[Th. 4.8]{massivemimobook}, which motivates to select the precoding vectors as scaled versions of the combining vectors $\vect{w}_{jk} = \frac{\vect{v}_{jk}}{ \sqrt{\mathbb{E}\{|| \vect{v}_{jk}||^2\}}}$
where the scaling factor is chosen to satisfy the precoding normalization constraint $\mathbb{E}\{||\vect{w}_{jk}||^2\} = 1$.
By selecting $\vect{v}_{jk}$ according to one of the UL combining schemes described earlier, the
corresponding precoding scheme is obtained. 

\textcolor{red}{The expectations in \eqref{eq:downlink-instant-SINR} can be computed for any arbitrary precoding scheme by using Monte Carlo simulations. However, similar to \cite[Cor. 4.5]{massivemimobook}, we can obtain the closed-form expressions when using MR precoding, as described in the following corollary.}

\begin{corollary}\label{theorem:DL-SE_MR}
If MR precoding is used with $\vect{w}_{jk} = \frac{\widehat{\vect{g}}_{jk}}{ \sqrt{\mathbb{E}\{|| \widehat{\vect{g}}_{jk}||^2\}}}$, the expectations in \eqref{eq:downlink-instant-SINR} become $$| \mathbb{E} \{\vect{w}_{jk}^{\Htran} \vect{g}_{jk}^j \} |^2  = p_{jk}\tr \!\left( \left(\vect{u}_{jk}\vect{u}_{jk}^{\Htran}\right)\otimes\vect{\Phi}_{jk}^j \right),$$
and
\begin{align}
\mathbb{E} \{ | \vect{w}_{li}^{\Htran} {\vect{g}}_{jk}^l |^2 \} = \frac{ \tr \!\left( \left(\left(\vect{u}_{jk}\vect{u}_{jk}^{\Htran}\right)\otimes \vect{R}_{jk}^{l}\right) \left(\left(\vect{u}_{li}\vect{u}_{li}^{\Htran}\right)\otimes\vect{\Phi}_{li}^j \right)\right) \! }{ \tr \!\left( \left(\vect{u}_{li}\vect{u}_{li}^{\Htran}\right)\otimes\vect{\Phi}_{li}^l \right) \! }.
\end{align}
\end{corollary}

If the spreading signatures $\{\vect{u}_{jk}\}$ are selected from a set of mutually orthogonal vectors, then we can choose $\vect{W}_{jk}= \bar{\vect{w}}_{jk} \vect{u}_{jk}^{\Htran}$ where $ \bar{\vect{w}}_{jk} \in \mathbb{C}^{M}$ is the precoding vector associated to UE $k$ in cell $j$. 
Therefore, \eqref{eq_received_signal_DL0} reduces to
\begin{align} 
z_{jk} &= \vect{y}_{jk}^{\Htran}{\bf {u}}_{jk} = N\varsigma_{jk} (\vect{h}_{jk}^j)^{\Htran}\bar{\vect{w}}_{jk} \nonumber \\
&+ \sum\limits_{(l,i)\in\mathcal{C}_{jk}} N \varsigma_{li} (\vect{h}_{jk}^j)^{\Htran}\bar{\vect{w}}_{li}  + \vect{n}_{jk}^{\Htran} {\bf {u}}_{jk},
\end{align}
from which the effective SINR in \eqref{eq:downlink-instant-SINR} reads as
\begin{align}\label{eq:downlink-instant-SINR-orthogonal}
\gamma_{jk}^{\rm {dl}} 
& =  \frac{ \rho_{jk}| \mathbb{E} \{\vect{w}_{jk}^{\Htran} \vect{h}_{jk}^j \} |^2  }
{ 
\!\sum\limits_{(l,i)\in \mathcal{C}_{jk}}  \rho_{li} \mathbb{E} \{ | \bar{\vect{w}_{li}}^{\Htran} {\vect{h}}_{jk}^l |^2 \} - \rho_{jk}| {\mathbb{E}} \{ \bar{\vect{w}}_{jk}^{\Htran} {\vect{h}}_{jk}^j \} |^2
+  \frac{\sigma_{\rm{dl}}^{2}}{N}
}
\end{align}
{\color{black} where the noise power is reduced by a factor $N$ compared to classical mMIMO (see \cite[Th. 4.6]{massivemimobook})}. 
{If MR precoding is used with $\bar{\vect{w}}_{jk} = {\widehat{\vect{h}}_{jk}}/{ \sqrt{\mathbb{E}\{|| \widehat{\vect{h}}_{jk}||^2\}}} $, then \eqref{eq:downlink-instant-SINR-orthogonal} reduces to \eqref{eq:downlink-rate-expression-forgetbound-MR-orthogonal} (see above).

Unlike with mMIMO (e.g., \cite[Cor. 4.7]{massivemimobook}), the strength of coherent and non-coherent interference terms is determined by how similar the spatial correlation matrices $\vect{R}_{li}^{l}$ with $(l,i)\in \mathcal{C}_{jk}$ and $(l,i) \in \{\Pu_{jk}\cap\mathcal{C}_{jk}  \setminus  (j,k) \}$ are to $\vect{R}_{jk}^{l}$. By assigning orthogonal spreading signatures to the UEs with similar channel conditions, the SE can be higher than with mMIMO. We notice also that a $N-$ fold reduction of the noise term is achieved.
}

\begin{table*}[t]
\renewcommand{\arraystretch}{1.}
\centering
\caption{Network parameters}
\label{table:system_parameters_running_example}
\begin{tabular}{|c|c|}
\hline \bfseries $\!\!\!\!\!$ Parameter $\!\!\!\!\!$ & \bfseries Value\\
\hline\hline

      Cell size &  $250$\,m $ \times\,  250$\,m \\
UL noise power& $\sigma^2 = -94$ \\
UL and DL transmit powers & $p_{jk}= \rho_{jk}= 20$\,dBm \\
%

Samples per coherence block & $\tau_c = 200$ \\


Distance between UE $i$ in cell $l$ and BS~$j$ & $d_{li}^{\,j}$ \\

\begin{tabular}{@{}c@{}} Large-scale fading coefficient for \\the channel between UE $i$ in cell $l$ and BS~$j$\end{tabular}
& $\beta_{li}^{j} =  -148.1 - 37.6 \, \log_{10} \left( \frac{d_{li}^{j}}{1\,\textrm{km}} \right) + F_{li}^{j}$\,dB\\
 
Shadow fading between UE $i$ in cell $l$ and BS~$j$ & $F_{li}^{j} \sim \mathcal{N}(0,10)$ \\




\hline
\end{tabular}\vspace{0.7cm}
\end{table*}


\section{Numerical analysis for the case study: Single-cell with two UEs}\label{Sec:NumericalAnalysis}
To quantify the potential benefits of code-domain NOMA in mMIMO, we begin by considering the simple case study of Section II with $L=1$, $M=64$, and $K=2$, and numerically evaluate the SE for the practical setup described in Table \ref{table:system_parameters_running_example}. For brevity, the analysis is carried out in the UL and MR and MMSE combining using MMSE channel estimation are considered. When NOMA is employed, we assume that orthogonal codes of length $N=2$ are assigned to the two UEs. The two practical channel models described in Section~\ref{Channel_modeling} are used.

\subsection{Is NOMA needed?}
Similar to Fig.~\ref{fig:fig_SE_toy_example}, we investigate the SE behavior with respect to the UEs' locations. We fix the nominal azimuth angle of one UE at $30^{\circ}$ while we let the nominal azimuth angle of the second one vary from $-90^{\circ}$ to $90^{\circ}$. Following the setup in Fig.~\ref{fig:fig_SE_toy_example}, we impose that the average channel gain per antenna stays the same, i.e., $\beta_{11}^1=\beta_{12}^1$. Fig. \ref{SE_funcOfAoI_1cell2UEs_UL} shows the UL SE of UE 1 with classical mMIMO and mMIMO-NOMA for the 2D and 3D models. \textcolor{blue}{With the NOMA scheme, N-MMSE and N-MR are exactly the same since $N=2$ and thus no interference is present---this is why only the N-MMSE curve is reported.} Both channel models are considered with a relatively small ASD of $\Delta = 2^{\circ}$. We observe that classical mMIMO gives higher SE than NOMA in both 2D and 3D models for most of the angles of the interfering UE. Different results are obtained for the case in which the two UEs have very similar angles. This is a challenging setup characterized by unfavorable propagation, wherein NOMA can bring some benefit. 

For the 2D model, Fig.~\ref{SE_funcOfAoI_1cell2UEs_2D_UL} shows that  MMSE largely outperforms NOMA even in this poor favorable propagation condition. This is because MMSE is a sufficiently powerful scheme to reject the interference even when the UEs are very close in space. However, we notice that this is achieved at the cost of a higher computational complexity than with MR \cite{massivemimobook} since the complexity scales as $M^3$. Fig. \ref{SE_funcOfAoI_1cell2UEs_UL} also shows that NOMA can provide some gain compared to MR, without any increase in complexity. 

For the 3D model, Fig.~\ref{SE_funcOfAoI_1cell2UEs_3D_UL} reveals that, when the UEs are close in space, NOMA provides the highest SE irrespective of the combining scheme used with mMIMO. This is because the planar array has a smaller spatial resolution, that reduces the spatial interference rejection capabilities of mMIMO and opens the door for complementing it with NOMA.

\subsection{A look at the favorable propagation conditions}

 \begin{figure} 
        \centering \vspace{-0.7cm}
        \begin{subfigure}[b]{\linewidth} \centering 
                \includegraphics[width=1\linewidth]{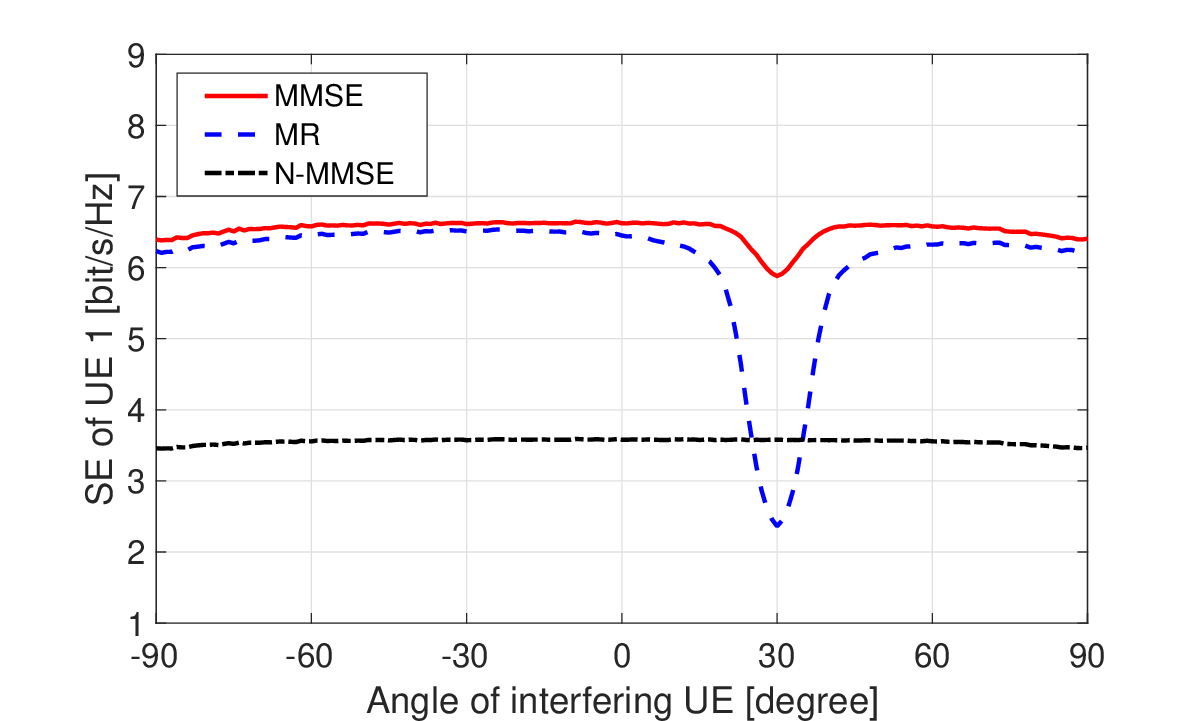}  
                \caption{UL with the 2D channel model} 
                \label{SE_funcOfAoI_1cell2UEs_2D_UL}
        \end{subfigure} 
        \begin{subfigure}[b]{\linewidth} \centering
                \includegraphics[width=1\linewidth]{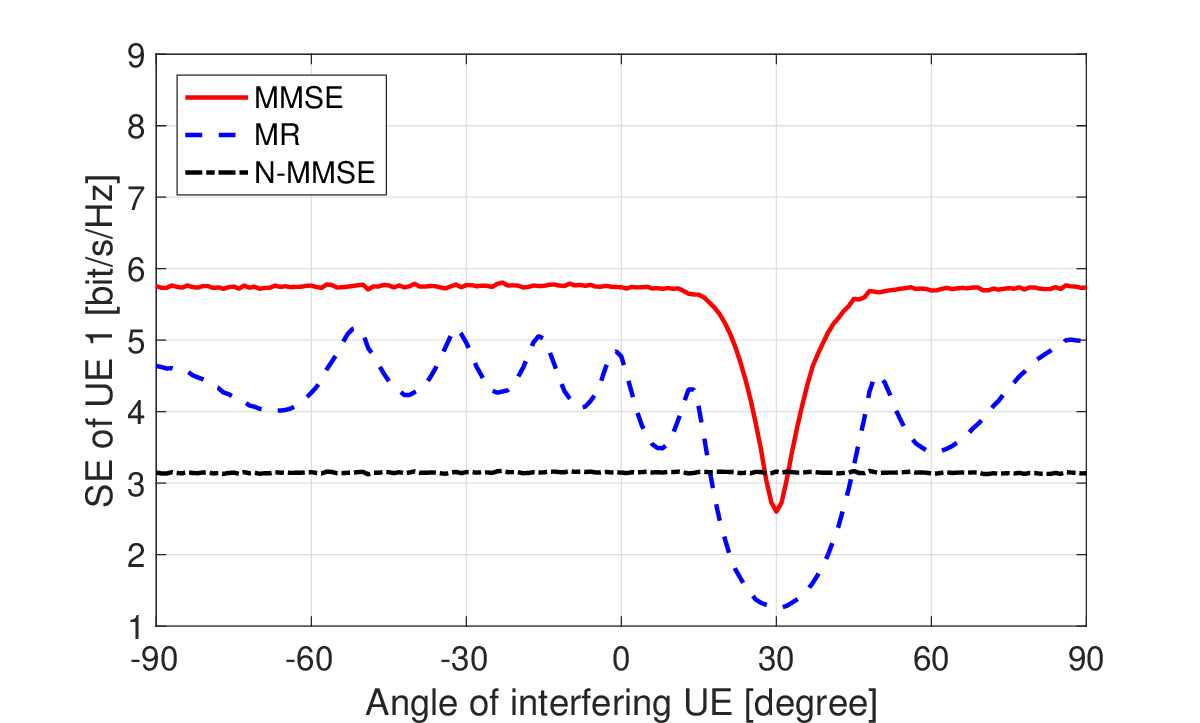} 
                \caption{UL with the 3D channel model}  
                \label{SE_funcOfAoI_1cell2UEs_3D_UL} 
        \end{subfigure}  
\vspace{-0.25cm}
\caption{SE of UE 1 in a single-cell two-user setup with $\Delta=2^{\circ}$ and $M = 64$ with mMIMO and mMIMO-NOMA for $N=2$ as a function of the azimuth angle of the interfering UE. The nominal azimuth angle of the desired UE is fixed at $30^{\circ}$. The 2D (Fig. \ref{SE_funcOfAoI_1cell2UEs_2D_UL}) and 3D (Fig. \ref{SE_funcOfAoI_1cell2UEs_3D_UL}) channel models described in Section~\ref{Channel_modeling} are considered.}\label{SE_funcOfAoI_1cell2UEs_UL}
\end{figure}

To better understand the above results, Fig.~\ref{fig:figVar_funcOfAoI} shows the variance
\setcounter{equation}{30}
\begin{align} 
\delta_{1,12}^1&=\mathbb{V}  \left \{ 
\frac{(\vect{h}_{11}^1)^{\Htran}\vect{h}_{12}^{1}}{\sqrt{\mathbb{E} \{ \| \vect{h}_{11}^1 \|^2 \} \mathbb{E} \{ \| \vect{h}_{12}^{1}  \|^2 \} } }
\right \}= \frac{\tr \left( \vect{R}_{11}^1 \vect{R}_{12}^1 \right) }
 { M^2\beta_{11}^1 \beta_{12}^1}\label{eq:variance-favorable-propagation}
\end{align}
of the two UEs for 2D and 3D models in the same setup of Fig.~\ref{SE_funcOfAoI_1cell2UEs_UL}. The variance is quantitatively measuring the level of favorable propagation \cite[Eq. (2.19)]{massivemimobook}. It takes values in the interval $\delta_{1,12}^1\in [0,1]$, where smaller values represent a higher level of favorable propagation. Specifically, $\delta_{1,12}^1=1$ if $\vect{R}_{11}^1$ and $\vect{R}_{12}^1$ are rank one and have the same dominant eigenvector.

In contrast, $\delta_{1,12}^1=0$ if the correlation matrices $\vect{R}_{11}^1$ and $\vect{R}_{12}^1$ are orthogonal, i.e., $\tr \left( \vect{R}_{11}^1 \vect{R}_{12}^1 \right) = 0$, which is a special case of linearly independent correlation matrices. Note that full orthogonality is unlikely to appear in practice \cite{Sanguinetti2019a}. 

\begin{figure}[t!]
\centering\vspace{-0.7cm}
\includegraphics[unit=1mm,width=1\linewidth]{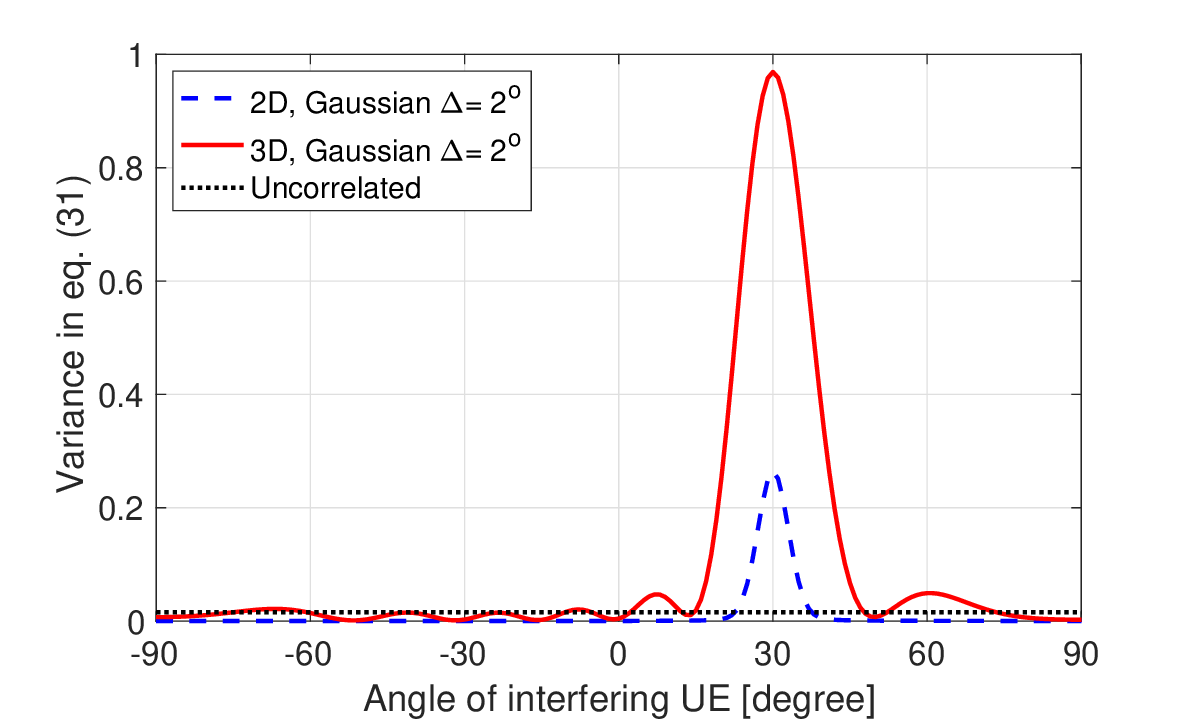}
\caption{{Behaviour of the variance defined in eq. \eqref{eq:variance-favorable-propagation}, for the same setup of Fig.~\ref{SE_funcOfAoI_1cell2UEs_UL}, with respect to the change of azimuth angle of the interfering UE. Uncorrelated fading is also reported for comparison with 2D and 3D channel models, described in Section~\ref{Channel_modeling}.}}
\label{fig:figVar_funcOfAoI}
\end{figure}

The variance in \eqref{eq:variance-favorable-propagation} equals $1/M^2$ for uncorrelated fading channels. However, Fig.~\ref{fig:figVar_funcOfAoI} shows that the values of \eqref{eq:variance-favorable-propagation} changes with angles when considering the 2D and 3D channel models. It achieves its maximum value at $30^{\circ}$ for both models, which coincides with the angle giving the lowest SE values in Fig.~\ref{SE_funcOfAoI_1cell2UEs_UL}. With the 2D model, the peak variance is relatively small ($\approx 0.25$), leading to comparatively good favorable propagation conditions. This justifies why classical mMIMO performs fairly well in the setup of Fig.~\ref{SE_funcOfAoI_1cell2UEs_UL}. On the other hand, the variance is substantially larger ($\approx 0.95$) with the 3D model. This is because both horizontal and vertical spatial resolutions of the $8 \times 8$ array is only given by $8$ antennas. Therefore, separation of the UEs in any of the two domains cannot be achieved. {\color{black}Hence, the two UEs cause much interference to each other, and thus the SE of mMIMO deteriorates, especially with MR. As shown in Fig.~\ref{SE_funcOfAoI_1cell2UEs_UL}, this issue can be solved with NOMA by assigning orthogonal spreading signatures to the UEs with similar channel conditions. A natural question is thus how to group the UEs in a cell into groups that offer favorable propagation conditions. This problem is addressed next.}

\section{UE Grouping}\label{Sec:UEgrouping}

The concept of grouping UEs in mMIMO based on their spatial correlation matrices was introduced in \cite{Huh2012a}, but for the purpose of orthogonal time-frequency scheduling when the UEs in each group have identical low-rank spatial correlation matrices. {\color{blue}Inspired by \cite{Huh2012a}, the vast majority of UE scheduling algorithms (e.g., \cite{Zhu19} and references therein) rely on the sparsity of channels (i.e., rank-deficient correlation matrices). However, channel measurements for mMIMO systems operating in sub$-6$ GHz bands have recently shown that the spatial correlation matrices may have high rank, with a mix of several weak and a few strong eigendirections \cite{Gao2015c,Flordelis2018}, and vary even between closely spaced UEs; see also \cite[Section III.C]{Sanguinetti2019a} for a discussion on the main properties of practical spatial correlation matrices. This implies that one cannot separate UEs into groups with orthogonal spatial correlation matrices to guarantee favorable propagation conditions, or expect UEs in the same group to have identical statistics.} In other words, the grouping of UEs is highly non-trivial and will be addressed in this section.
To this end, we first define the notion of dominant eigenspaces to capture the eigenspace that contains most of the energy of each correlation matrix.

{\color{black}
\LinesNotNumbered
\begin{algorithm}
\SetAlgoLined
\KwInput{$\{\vect{R}_{jk}^j; k=1,\ldots,K\}$, $G$, and $p$}
\KwOutput{$\{\bar{\vect U}_1,\ldots,\bar{\vect U}_g\}$, $\{{\vect U}_1,\ldots,{\vect U}_K\}$, and $\{C_1,\ldots,C_G\}$} 
 
 \tcc{Compute the $p-$dominant eigenspaces of all UEs}  
 
  \For{$k=1,\ldots,K$}{${\vect U}_k \gets \eig_p(\vect{R}_{jk}^j)$}
 
  \tcc{Initialization}
 
   $t \gets 0$ \\
  \tcc{Select the initial group indicies}   
  \For{$g=1,\ldots,G$}{
  \tcc{Set a random UE as the group center}
    \hspace{2mm} $C_g^t \gets \{i_g\}$, $i_g$ $\in \{1,\ldots,K\}\setminus \cup_{g=2}^{G}C_{g-1}^t$ \\
 \tcc{Compute the group eigenspaces}
   \hspace{2mm} $\bar{\vect U}_g \gets {\vect U}_{i_g}$}
   
 \tcc{Iteratively group the $p-$ dominating eigenspaces}  
 \While{$\{C_1,\ldots,C_G\}\ne \{C_1^{t-1},\ldots,C_G^{t-1}\}$}{
  $t \gets t +1$\\
\tcc{Prepare to save the new UE indices of each group} 
 
  \For{$g=1,\ldots,G$}{$C^t_g \gets \emptyset$}
 
\tcc{Assign UEs to the nearest group} 
  \For{$k=1,\ldots,K$} {$g_k^\star \gets \arg  \underset{g}{\min}  \; \mathrm{d}_\textrm{C}(\bar{\vect U}_g,{\vect U}_k) $\\$C_{g_k^\star}^t \gets C_{g_k^\star}^t \cup {k} $}
  
   \tcc{Recompute the group mean} 
\For{$g=1,\ldots,G$}{$\bar{\vect U}_g \gets \eig_p\left(\sum_{k\in C_g^t}{\vect U}_{k}{\vect U}_{k}^{\Htran}\right)$\\$C_g \gets C_g^t$}
 }
 \caption{$k-$means algorithm for cell $j$.}
\end{algorithm}
}

\begin{definition}[$p$-Dominant eigenspace]\label{def:pdom-eigspace}
\index{dominant eigenspace}
Let $\vect{A}\in\mathbb{C}^{M\times M}$ be a Hermitian matrix with eigenvalue decomposition $\vect{A} = \vect{U}\vect{D}\vect{U}^{\Htran}$. The $p$-dominant eigenspace $\eig_p(\vect{A}) = \left[ \vect{u}_1 \ldots \vect{u}_p \right]$ is the (tall) unitary matrix composed of the $p$ eigenvectors belonging to its $p$ largest eigenvalues.
\end{definition}

The problem is how to group the UEs in a cell such that the $p-$dominating eigenspaces of the (possibly full-rank) correlation matrices of the UEs in each group are similar and different from the correlation matrices of other groups. A similarity score metric for measuring the difference between two eigenspaces is needed. A possible choice is given by the \textit{chordal distance}. 

\begin{definition}[Chordal distance]\label{Chordal_Distance}
\index{chordal distance}
The \emph{chordal distance} $\mathrm{d}_\textrm{C}(\vect{A},\vect{B})$ between two matrices $\vect{A}$ and  $\vect{B}$ is defined as 
\begin{equation}\label{eq:chordal-distance}
\mathrm{d}_\textrm{C}(\vect{A},\vect{B}) = \|\vect{A}\vect{A}^{\Htran}-\vect{B}\vect{B}^{\Htran}\|^2_F.
\end{equation}
\end{definition}

{\color{black}For two (tall) unitary matrices $\vect{A},\vect{B}\in \mathbb{C}^{M\times p}$, the chordal distance takes the form
\begin{align}\notag
\mathrm{d}_\textrm{C}(\vect{A},\vect{B}) & = \|\vect{A}\vect{A}^{\Htran}-\vect{B}\vect{B}^{\Htran}\|^2_F \\ \notag
&= \tr((\vect{A}\vect{A}^{\Htran}-\vect{B}\vect{B}^{\Htran})(\vect{A}\vect{A}^{\Htran}-\vect{B}\vect{B}^{\Htran})^{\Htran})\\\notag& = \tr(\vect{A}\vect{A}^{\Htran}+\vect{B}\vect{B}^{\Htran}-2\vect{A}\vect{A}^{\Htran}\vect{B}\vect{B}^{\Htran})\\& = 2p-2\sum_{i=1}^p\sum_{j=1}^p|\vect{a}_i^{\Htran}\vect{b}_j|^2\label{eq:chordal-distance-1}
\end{align}
where $\vect{a}_k$ and $\vect{b}_k$ denotes the $k$th column of $\bf A$ and $\bf 
B$, respectively.
The chordal distance can be interpreted as the number of dimensions of the subspace that can be reached by a linear combination of the column vectors of only one of the two matrices. For example, if $\vect{A} = \vect{B}$, we have $\mathrm{d}_\textrm{C}(\vect{A},\vect{B}) = 0$. Although each matrix individually spans $p$ dimensions, all of them can be reached through a linear combination of the column vectors of $\vect{A}$ and $\vect{B}$. On the other hand, for $\vect{A}^{\Htran} \vect{B} = {\bf 0}_{p\times p}$, we have $\mathrm{d}_\textrm{C}(\vect{A},\vect{B}) = 2p$ because each matrix spans a $p-$dimensional space which cannot be reached through a linear combination of the column vectors of the other matrix.}

Several solutions exist in the literature to form groups on the basis of similarity scores \cite{Huh2012a,Ko2012,Kuda2019NOMAaided,Zhu19}. Among those, we adopt the $k-$means algorithm, which is widely used and operates as follows. For any cell $j$, $k-$means takes as inputs the set of intra-cell spatial correlation matrices $\{\vect{R}_{jk}^j; k=1,\ldots,K\}$, the desired number of groups $G$, and the desired number of dominant eigenspace dimensions per group $p$. The output is a set of $G$ tall unitary matrices $\{\bar{\vect U}_g\in \mathbb{C}^{M\times p}:g=1,\ldots,G\}$, representing the center (or mean) of each group, and the sets $\{C_g:g = 1,\ldots,G\}$, where $C_g$ denotes the index set of UEs belonging to group $g$. The pseudo-code provided in Algorithm 1 describes how the algorithm works. {\color{blue}Notice that the use of chordal distance in Algorithm 1 has three advantages: $i$) it can be used to measure the difference of possibly full-rank correlation matrices; $ii$) it reduces the computational complexity since only the $p-$dominant eigenspaces (with $p\ll M$) of each UE are used; $iii$) it can be applied with the $k-$means algorithm since it is an Euclidean distance. If the latter condition was not satisfied, solutions can be found using the $k-$medoid algorithm, which has higher complexity. The $k-$medoid algorithm is for example used in \cite{Zhu19} based on the normalized channel correlation factor $\frac{(\vect{h}_{jk}^j)^{\Htran}\vect{h}_{ji}^{j}}{ \| \vect{h}_{jk}^j \|  \| \vect{h}_{ji}^{j}  \|  }$,
which requires perfect channel state information. Heuristic solutions can also be found using greedy algorithms (e.g. \cite{Kuda2019NOMAaided}).}

\begin{figure}[t!]
\centering\vspace{-1cm}
\includegraphics[unit=1mm,width=1\linewidth]{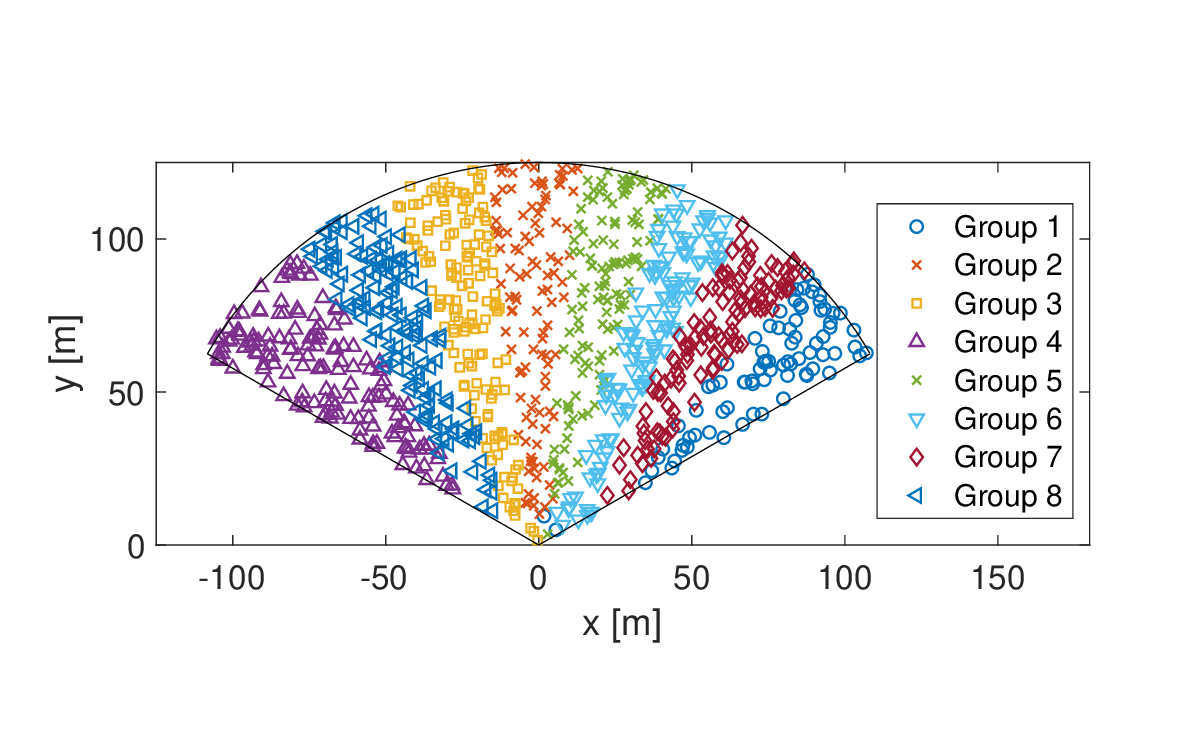}\vspace{-0.5cm}
\caption{{Resulting association of UE positions to groups from the $k-$means algorithm with $G = 8$, $p = 6$ and a 3D one-ring channel model with a planar quadratic $8\times 8$-antenna array.}}\vspace{-0.2cm}
\label{fig:grouping}
\end{figure}

The $k-$means algorithm allows us to partition a cell into geographical regions, which are characterized by correlation matrices spanning almost orthogonal dominant eigenspaces. {\color{black}This concept can be seen as a three-dimensional extension of the traditional cell sectorization. While the latter is static and fixed at the time of the antenna deployment, the former covariance-based clustering algorithm adapts dynamically to the UE locations and the propagation environment.} The algorithm can be applied 'offline' to a very larger number of correlation matrices, which have been recorded over time to find static, but environment dependent, group spaces. Only the association of UEs to groups needs to be computed at the run-time. An example of offline grouping is provided in Fig.~\ref{fig:grouping}, which shows the resulting association of 1000 UE positions to $G = 8$ groups of $p = 6$ dimensions under the 3D one-ring model for a planar quadratic $8\times 8$ antenna array with half-wavelength-spacing. The UEs are uniformly distributed over a $120^\circ$ sector with $125$\,m radius. {\color{black}Note that the algorithm has partitioned the cell into eight  azimuth bins while no separation is visible in the elevation dimension. This is because the horizontal angular
{\color{black}
\begin{algorithm}
\SetAlgoLined
\KwInput{$\{\vect{R}_{jk}^j; k=1,\ldots,K\}$, $G$, and $p$}
\KwOutput{$\{C'_1,\ldots,C'_G\}$}
 
\rm{ \textcolor{blue}{\textbf{Step 1} - Preliminary UEs grouping} } \\
 \tcc{ \textcolor{blue}{Run the $k-$means algorithm in Algorithm 1}}
	$\{\bar{\vect U}_1,\ldots,\bar{\vect U}_g\}$;   $\{{\vect U}_1,\ldots,{\vect U}_K\}$; $\{C_1,\ldots,C_G\}$
 
\rm{ \textcolor{blue}{\textbf{Step 2} - Assignment problem}}  
  
\tcc{Find the distances between UEs and group centers}
\For{$k=1,\ldots,K$}{
   		\For{$g=1,\ldots,G$}{$d_{g,k} \gets \mathrm{d}_\textrm{C}(\bar{\vect U}_g,{\vect U}_k) $}
 }
 
$\vect{D} = \{d_{g,k}; g=1,...,G, k =1,...,K\}$ \tcp{Obtain the distance matrix of size $G\times K$}
 
 \tcc{Formulate the square matrix (Hungarian matrix)}
$\vect{H}=\{d_{r,k}; r=1,...,K, k =1,...,K\} \gets \vect{D} \kron {\bf 1}_{N}$   \tcp{replicate $N-$times each row of $\mathrm{D}\in \mathbb{R}^{G\times K}$ to obtain $\vect{H}\in \mathbb{R}^{K\times K}$}
\tcc{Initialize the group slot indices}
\For {$r=1,\dots,K$}{$i_r \gets \emptyset$}
\tcc{Implement the Hungarian method}
\For {$r=1,\dots,K$}{
\tcc{Find a unique column $k$ of each row with the smallest value}

\If {$\{k^{\star} \gets \underset{k}{\arg \min \, } d_{r,k} \} \cup \{d_{r,k^{\star}} \neq d_{r',k^{\star}}\},$ \text{where} $r' \in \{1,\dots,K\} \backslash \{r\}$ } 
{ 
	\tcc{Update the index of the selected column of each row}
	$i_r \gets {k^{\star}}$	}
}
\tcc{Map the index of the selected column of each row to the new respective group slot}
\For{$g=1,\dots,G$}{
$C'_g \gets \{i_{g(N-1)+1},\dots,i_{gN}\} $}
 \caption{UE grouping algorithm for cell $j$.}
\end{algorithm}
}
spread dominates the vertical angular spread in the chosen scenario. For a smaller cell radius, a higher mounting height, a larger vertical antenna spacing, also groups in the elevation dimension can appear.}

 \begin{figure} 
        \centering \vspace{-0.5cm}
        \begin{subfigure}[b]{\linewidth} \centering
                \includegraphics[width=1\linewidth]{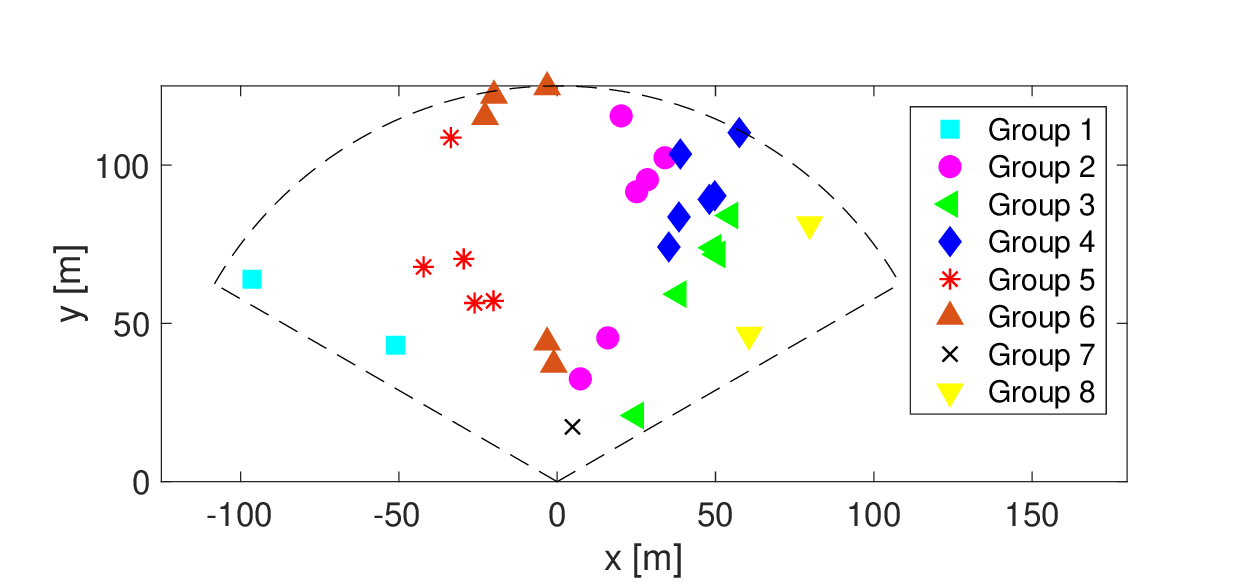} 
                \caption{Application of Algorithm 1} 
        \end{subfigure} 
        \begin{subfigure}[b]{\linewidth} \centering 
                \includegraphics[width=1\linewidth]{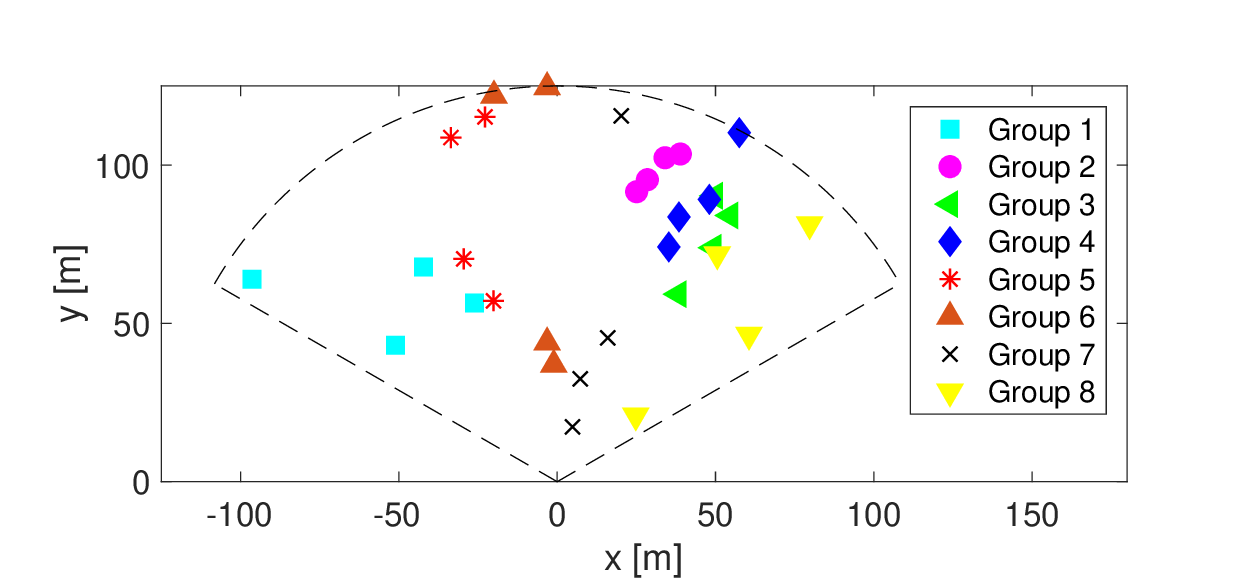} 
                \caption{Application of Algorithm 2}   
        \end{subfigure} 
\caption{Resulting association of $K=32$ UE positions to $G=8$ groups with Algorithm 1 and Algorithm 2. The latter allows to assign exactly $N=4$ UEs to each group.}\label{fig:Algorithm2}
\end{figure}

If the number of active UEs is not very large or UEs are located close to each other, Algorithm 1 may provide some groups that are empty while others are overloaded. {\color{black}To solve this issue, a further step in the $k-$means algorithm is needed, which assigns exactly $N$ UEs to each group while minimizing the sum of the chordal distance pairs. This can be achieved by employing the Hungarian method \cite{kuhn1955hungarian}, which is a combinatorial optimization algorithm that solves an assignment problem. \textcolor{blue}{This leads to Algorithm 2, which takes as input the output of Algorithm 1, which is represented by the $G$ unitary matrices $\{\bar{\vect U}_g\in \mathbb{C}^{M\times p}:g=1,\ldots,G\}$, representing the center of each group, the sets $\{C_g:g = 1,\ldots,G\}$, and the $K$ tall unitary matrices $\{\bar{\vect U}_k\in \mathbb{C}^{M\times p}:k=1,\ldots,K\}$. }The output returns the set $\{C_g':g = 1,\ldots,G\}$, where $C_g'$ denotes the updated index set of those UEs being reallocated to group $g$. The sets of matrices $\{\bar{\vect U}_g, g=1,\ldots,G\}$ and $\{\bar{\vect U}_k, k=1,\ldots,K\}$ are used to obtain the matrix $\vect{D} \in \mathbb{R}^{G\times K}$, whose generic element $d_{g,k}$ represents the distance between UE $k$ and the center of group $g$. The distance matrix $\vect{D}$ is then used to compute the square Hungarian matrix $\vect{H} \in \mathbb{R}^{K\times K} $. This is done through the following operation $\vect{H}= \vect{D} \kron {\bf 1}_N \in \mathbb{R}^{K\times K}$, which simply replicates $N$ times the $G$ rows of $\vect{D}$.\footnote{Notice that this step is needed because the Hungarian method works with square matrices. The $K-NG$ extra rows of $\vect{H}$ could also be made of all zeros without changing the output of the algorithm.} 
The algorithm proceeds by finding the minimum chordal distance (cost) when assigning UEs to groups based on cost, and such that each UE must be assigned to a different group. As mentioned above, the key of Algorithm 2 is that it assigns exactly $N$ UEs to each group such that the $N-$length spreading sequences can be efficiently used within each group. An example is provided in Fig.~\ref{fig:Algorithm2} for the same setup of Fig.~\ref{fig:grouping} but with $K=32$ UE positions. The resulting association to $G = 8$ groups is shown with both Algorithm 1 and Algorithm 2. Only the latter allows to assign exactly the same number of UEs to each group, which is in this case $N=4$. To the best of our knowledge, there exists no other UE-grouping algorithm in the literature that performs such operation.
}

\begin{remark} UE-grouping is a widely investigated topic in multi-user wireless communications. There exist several schemes in the literature that differ in terms of underlying method (optimal, heuristic, greedy,...), similarity score metric, available information (instantaneous channel estimates, statistical knowledge,...), computational complexity, channel models and so forth. A fair comparison among the existing solutions is very hard and is out of the scope of this work since it would require a fine-tuning of all the specific solutions. We believe that the combination of Algorithm 1 and Algorithm 2 represents a good baseline scheme to perform UE-grouping in the context of code-domain NOMA, and quantifies the benefits that it can bring into mMIMO.
\end{remark}

\section{Performance evaluation}\label{Sec:PerformanceEvaluation}
{\color{blue} This section compares the performance of mMIMO with vs. without NOMA, and validates the benefits of the grouping algorithm. A network with $L = 4$ cells is considered. Each cell has an area of 250 m $\times$ 250 m. {\color{blue}We numerically evaluate the average sum SE per cell in the UL and DL, i.e.: 
\begin{equation} 
\mathsf{SE}_{j}^{\rm {ul}} = \sum\limits_{k=1}^K\mathsf{SE}_{jk}^{\rm {ul}} \quad \text{and} \quad \mathsf{SE}_{j}^{\rm {dl}} = \sum\limits_{k=1}^K\mathsf{SE}_{jk}^{\rm {dl}},
\end{equation}}
for the network setup defined in Table~\ref{table:system_parameters_running_example}. Each BS is located in the center of its cell, has $M$ antennas, and serves $K$ UEs. The analysis is carried out with both MR and MMSE combining schemes, using MMSE channel estimation. Based on results of Section V, only a 3D channel model with a $8 \times 8$ planar array and a relative small $\Delta=2^{\circ}$ is considered.} If not otherwise specified, we assume that $\tau_{\rm p}=K$ orthogonal pilot sequences are used for channel estimation. 
\vspace{-0.3cm}

 \subsection{How efficient is the UE grouping algorithm?}\label{sec:UEgrouping_Evaluation}
We begin by assessing the benefits of properly grouping the UEs with mMIMO-NOMA in the UL, with the two following typical scenarios:

 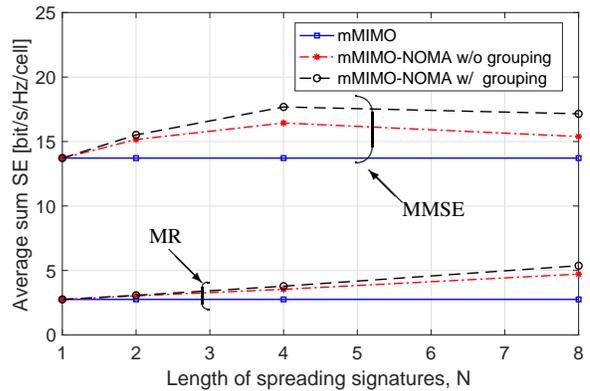
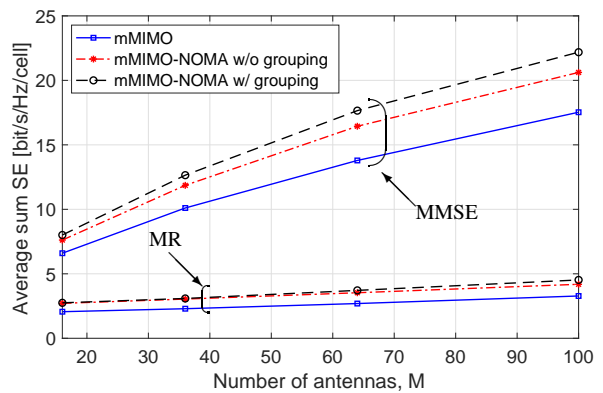
\begin{figure} 
        \centering \vspace{-0.7cm}
        \begin{subfigure}[b]{\linewidth} \centering 
               \begin{overpic}[unit=1mm,width=1\linewidth]{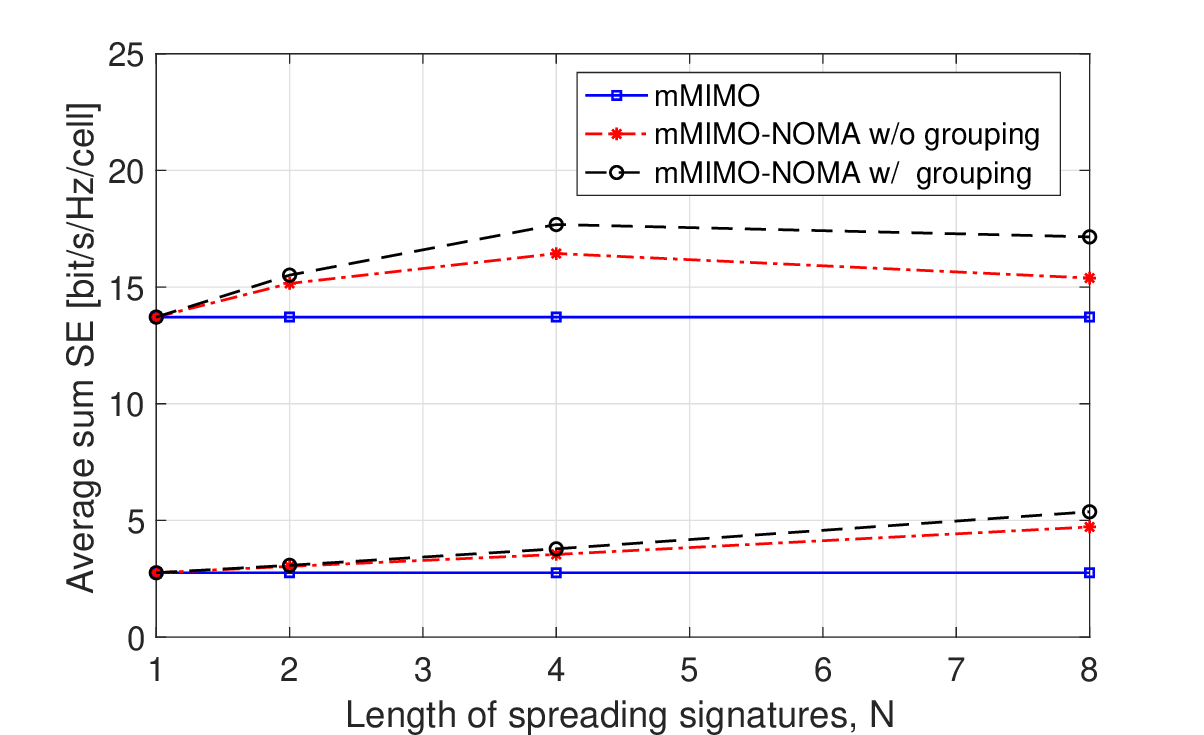}
	\put(64,26){\footnotesize MMSE}
		\put(64,28){\vector(-1, 1){5}}
	\put(57,39){\oval(5, 10)[r]}
		\put(26,22){\footnotesize MR}
	\put(29,21){\vector(1, -1){5}}
		\put(35,14){\oval(2, 4)[l]}
\end{overpic}
   \caption{Impact of spreading signature length} 
   \label{SE_funcOfN}
        \end{subfigure} 
  \begin{subfigure}[b]{\linewidth} \centering
                \begin{overpic}[unit=1mm,width=1\linewidth]{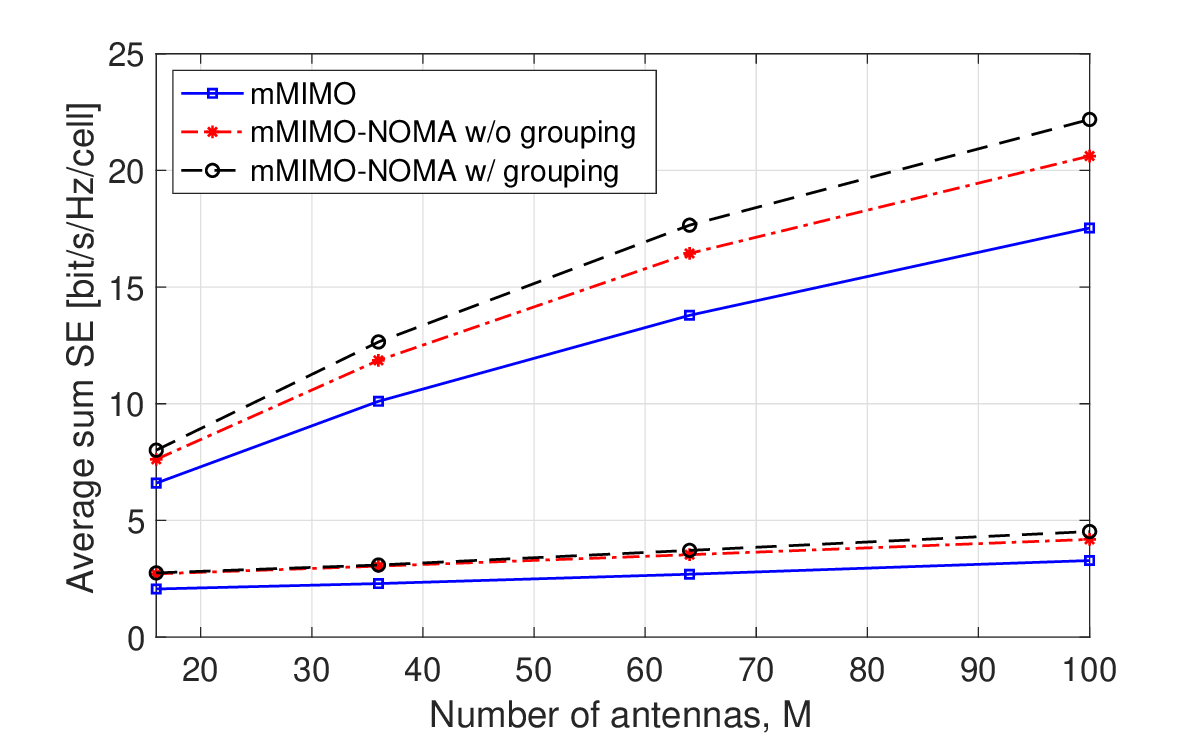}
		\put(66,26){\footnotesize MMSE}
		\put(66,28){\vector(-1, 1){5}}
	\put(59,39){\oval(5, 10)[r]}
		\put(26,22){\footnotesize MR}
	\put(29,21){\vector(1, -1){5}}
		\put(35,14){\oval(2, 4)[l]}
\end{overpic}
                \caption{Impact of the number BS antennas}  
      \label{SE_funcOfM}
        \end{subfigure}  
\vspace{-2mm}
\caption{{\color{blue}Average sum UL SE with mMIMO and mMIMO-NOMA as a function of the spreading signature length $N$ (Fig. \ref{SE_funcOfN}), and of the number of BS antennas $M$ (Fig. \ref{SE_funcOfM}) when $K=16$ UEs are uniformly distributed over a $30^{\circ}$ sector. mMIMO-NOMA is operated with no grouping vs. with a grouping algorithm (Algorithm 2).} Orthogonal spreading signatures are used.}\label{fig:UL_fixedK}
\vspace{-0.1cm}
\end{figure}       

\subsubsection{For a fixed number of UEs}
{\color{blue} Fig. \ref{SE_funcOfAoI_1cell2UEs_3D_UL} shows that SE is largely reduced when UEs are located within a $30^\circ$ sector. Therefore, we assume $K=16$ UEs uniformly and independently distributed over a $30^\circ$  sector (oriented as in Fig.~\ref{fig:grouping}), that is randomly located at a distance of $100$ m from the BS.} 
Fig.~\ref{fig:UL_fixedK} illustrates the average sum SE per cell, with classical mMIMO and mMIMO-NOMA. {\color{blue} With the latter scheme, the UE groups are formed either in a random way (i.e., without grouping) or through Algorithm 2 (i.e., with grouping). Sequences are orthogonal and randomly assigned to active UEs}. 
The impact of length of spreading signatures $N$ vs. number of BS antennas $M$ are shown in Fig.~\ref{SE_funcOfN} and Fig.~\ref{SE_funcOfM}, respectively. In Fig.~\ref{SE_funcOfN}, $N-$length signatures are assigned to the UEs in each group, implying that the number of formed groups is $G=K/N$. With $N=1$, there is no spreading and mMIMO-NOMA reduces to mMIMO. 

 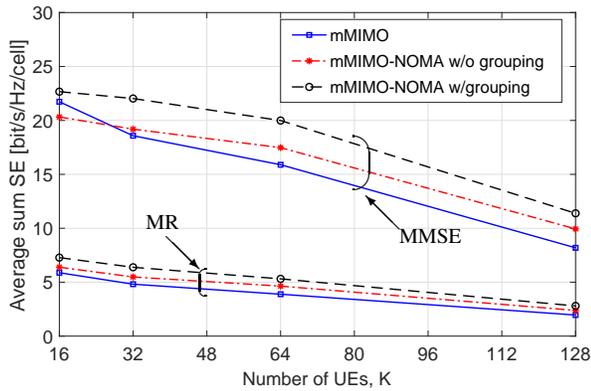
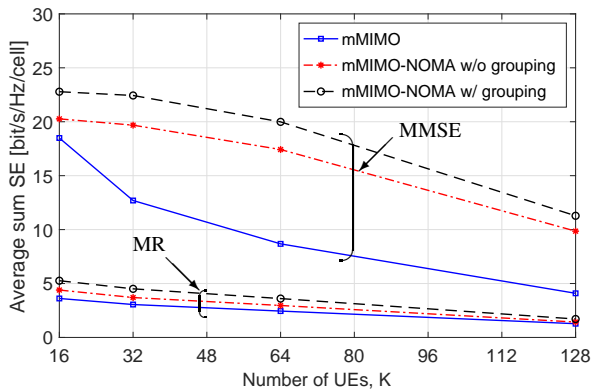
\begin{figure} 
        \centering \vspace{-0.7cm}
        \begin{subfigure}[b]{\linewidth} \centering 
               \begin{overpic}[unit=1mm,width=1\linewidth]{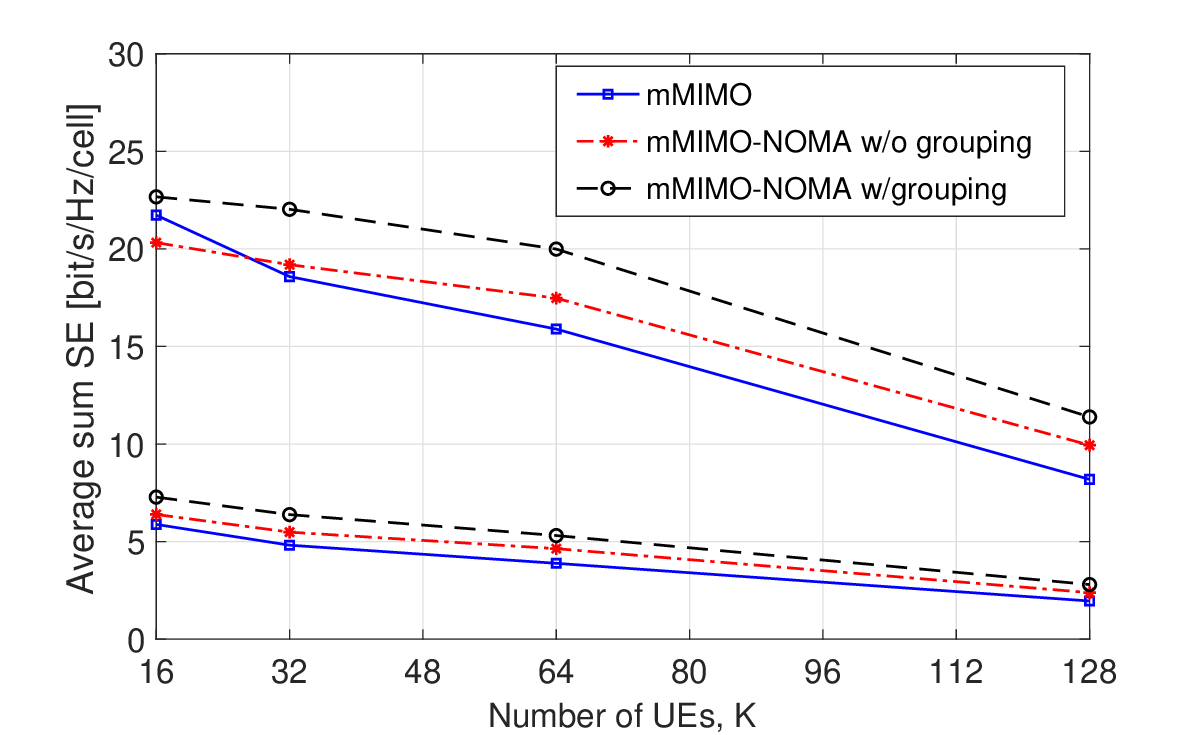}
	\put(64,22){\footnotesize MMSE}
		\put(64,24){\vector(-1, 1){5}}
	\put(57,34){\oval(5, 8)[r]}
		\put(26,24){\footnotesize MR}
	\put(29,23){\vector(1, -1){5}}
		\put(35,16){\oval(2, 4)[l]}
\end{overpic}
                \caption{UL transmission} 
         \label{fig:SE_UL}
        \end{subfigure} 
        \begin{subfigure}[b]{\linewidth} \centering
                \begin{overpic}[unit=1mm,width=1\linewidth]{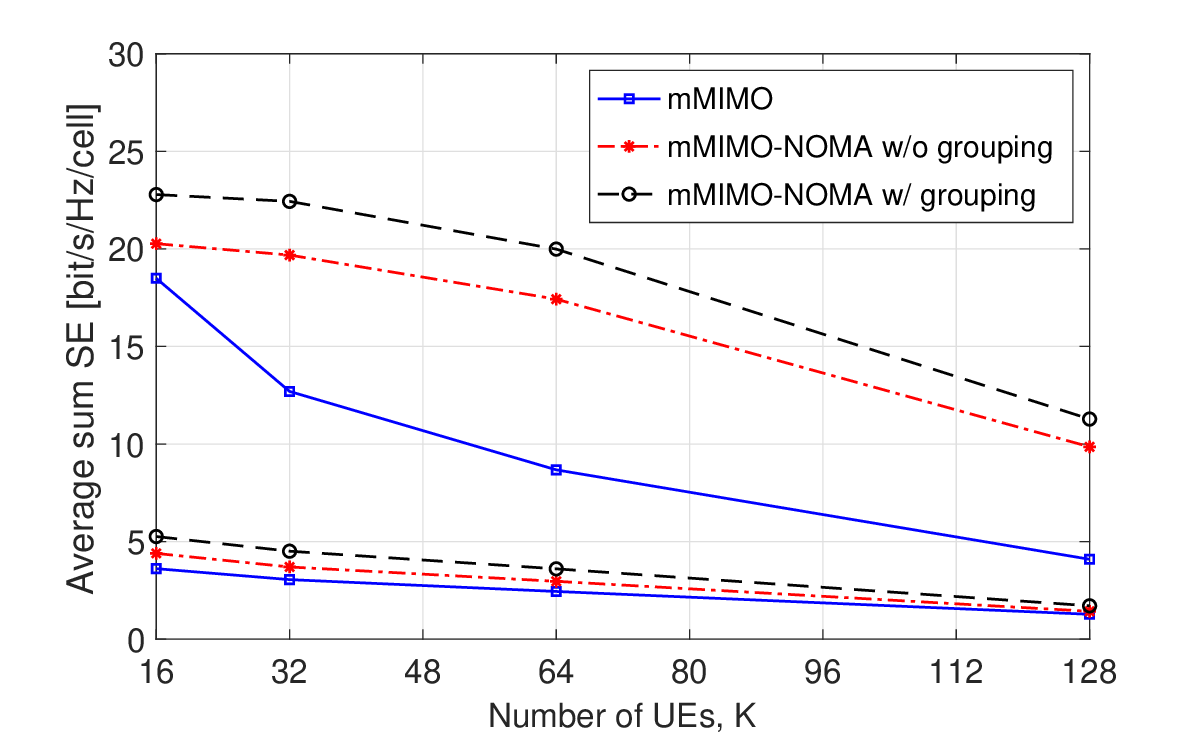}
		\put(55,29){\oval(4, 19)[r]}
	\put(64,38){\footnotesize MMSE}
	\put(63,38){\vector(-1, -1){5}}
		\put(24,21){\footnotesize MR}
	\put(29,20){\vector(1, -1){5}}
		\put(35,13){\oval(2, 4)[l]}
\end{overpic}
                \caption{DL transmission}  
        \label{fig:SE_DL}        
        \end{subfigure}  
\vspace{-2mm}
\caption{ {\color{blue} Average sum SE as a function of number of UEs $K$ with mMIMO and mMIMO-NOMA with no grouping vs. with a grouping algorithm (Algorithm 2).} UL and DL  transmissions are considered. Orthogonal spreading signatures are used.}\label{SE_funcOfK_multicell}
\vspace{-0.1cm}
\end{figure}

{\color{blue} Results of Fig.~\ref{fig:UL_fixedK} show that mMIMO-NOMA with Algorithm 2 achieves better performance than random grouping with both MR and MMSE combining, as shown, in particular, in Fig.~\ref{SE_funcOfN} for $N\geq 4$, and in Fig.~\ref{SE_funcOfM} for $M\geq 16$. Compared to mMIMO, both approaches of mMIMO-NOMA can provide higher gain, and the performance of MMSE is much greater than that of MR. This happens since MMSE combining has better interference cancellation capabilities. In summary, NOMA can bring some benefits compared to mMIMO also when spreading signatures are randomly assigned.} Better performance can be achieved if spreading sequences are assigned according to spatial correlation matrices. Results are in agreement with those of the case study (see Fig.~\ref{SE_funcOfAoI_1cell2UEs_UL}). In particular, Fig.~\ref{SE_funcOfM} confirms that there exists specific cases where NOMA can provide benefits even when $M \gg K$. Similar results can be obtained for the DL due to the UL-DL duality property, thus are skipped due to space limitations.

\subsubsection{Varying number of UEs}

We now consider the case in which the number of active UEs, $K$, in each cell increases. For an overall evaluation, we display the SE performance in both underloaded and overloaded regimes, i.e. $K$ ranges from 16 to 128, while the number of BS antennas is kept fixed at $M=64$. Similarly to Fig.~\ref{fig:UL_fixedK}, we assume that the UEs are located close to each other. Unlike Fig.~\ref{fig:UL_fixedK}, however, we assume that they are equally distributed in four distinct circle clusters with radius $r = 20$ m, that have $K/4$ UEs each, and are randomly deployed in each cell. This implies that the UEs are already grouped into $G=4$ groups per cell. Spreading signatures of length $N = K/4$  are assigned to the $K/4$ UEs in each group. Orthogonal spreading signatures are adopted. This might be a quite challenging setup for conventional mMIMO due to the insufficient spatial resolution of a planar BS array with 64 antennas. 

We compare classical mMIMO and mMIMO-NOMA with and without grouping-based signature assignment. {\color{blue} The average sum SE as a function of number of UEs $K$ is shown in the UL (Fig.~\ref{fig:SE_UL}) and DL (Fig.~\ref{fig:SE_DL}). With mMIMO-NOMA without grouping, the spreading sequences are randomly assigned to the UEs in the cell; this means that UEs in the same group can be assigned to the same spreading sequence.} The result shows that mMIMO-NOMA with proper assignment of sequences performs well in both UL and DL, in particular when using MMSE combining/precoding. mMIMO-NOMA with grouping achieves higher SE than classical mMIMO already with $K=16$, and the gap slightly increases as $K$ gets larger. With $K=32$, the SE gain is $20\%$ in the UL and $40\%$ in the DL. {\color{black} The constant gap between mMIMO-NOMA with grouping and classical mMIMO for both UL and DL {\color{blue} remains in the overloaded regime}, i.e. when $K>M, M =64$.} {\color{blue}The reason is that mMIMO-NOMA achieves a roughly constant sum SE as $K$ increases, while it reduces for classical mMIMO due to the lack of favorable propagation conditions. 
The SE reduction is larger in the DL than in the UL, which might be due to the suboptimality of MMSE precoding and equal DL power allocation.}

\begin{figure}[t!]
\centering\vspace{-0.7cm}
\begin{overpic}[unit=1mm,width=\linewidth]{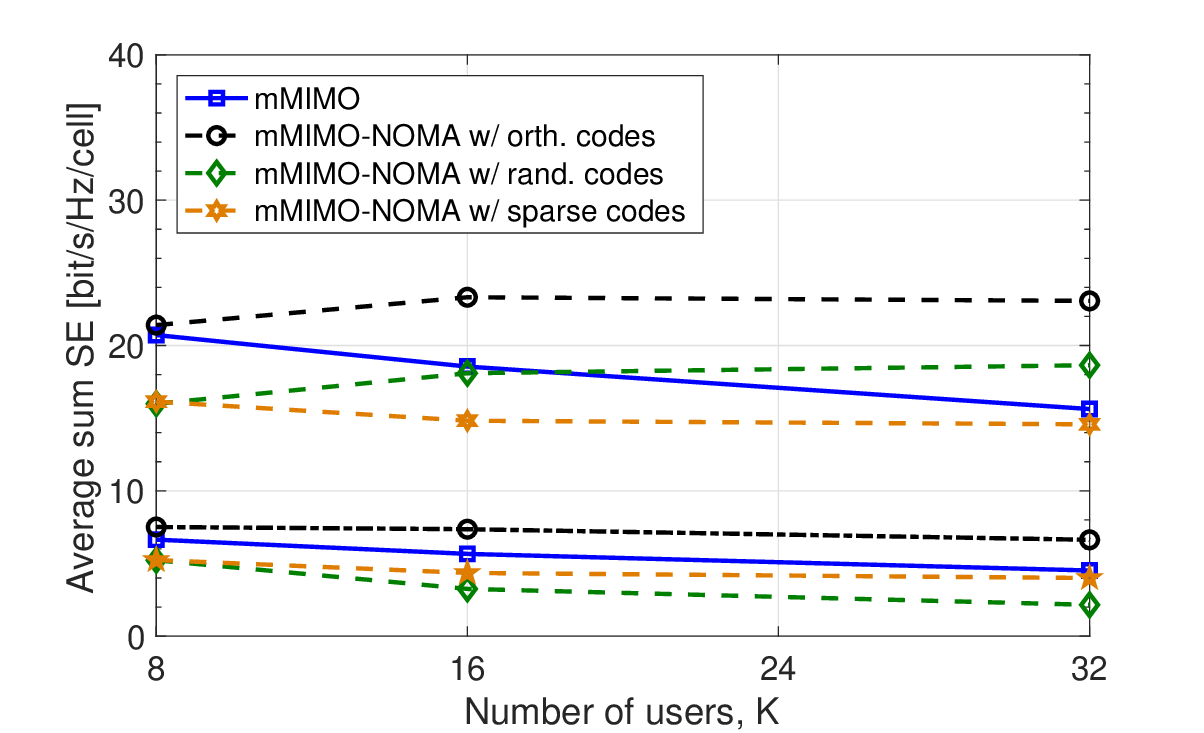}
		\put(55,31){\oval(5, 14)[r]}
	\put(60,43){\small MMSE}
	\put(63,42){\vector(-1, -1){5}}
		\put(26,23){\small MR}
	\put(29,22){\vector(1, -1){5}}
		\put(35,15){\oval(2, 8)[l]}
		\end{overpic}
\caption{Average sum UL SE as a function of number of UEs $K$ with mMIMO and mMIMO-NOMA for different types of spreading signatures of length $N= 4$.}\label{fig:UL_funcOfK_diffCodes}\vspace{-0.3cm}
\end{figure}
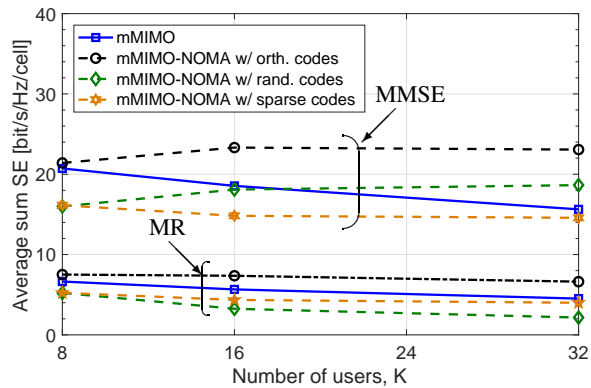

\vspace{-0.3cm}
\subsection{Which spreading signatures are more favorable?}
{\color{blue} We now compare the achievable SE with spreading signatures of length $N=4$, taken from either orthogonal, random, and sparse sets, as shown in Fig.~\ref{fig:UL_funcOfK_diffCodes}. In the random case, the $N$-length signatures are picked up from an assemble of $\{ \pm 1\}$, whereas in the sparse case, low-density signatures are used, having only one non-zero value randomly distributed within the $N$-length signature \cite{ferrante2015,LeTWC2018}.} Herein Fig.~\ref{fig:UL_funcOfK_diffCodes} shows the sum UL SE as a function of number of UEs in the same setup of Fig.~\ref{SE_funcOfK_multicell}. We notice that orthogonal signatures  give the highest performance with both MR and MMSE combining. {\color{blue} While mMIMO-NOMA with orthogonal codes has better performance for $K\ge 8$, mMIMO-NOMA with random codes might provide some gain compared to mMIMO for $K\ge 32$. This is because the probability that a given group of UEs is closely located in space increases as $K$ becomes larger. Interestingly, mMIMO-NOMA with MR outperforms mMIMO only when orthogonal codes are used; this is because MR cannot deal with the extra interference originating from the non-orthogonality of random and sparse codes.} \textcolor{red}{As in the case of Fig.~\ref{fig:UL_fixedK}, similar results are obtained for the DL, and thus omitted due to space limitations.}

\subsection{Impact of channel estimation quality}
The spatial interference rejection capabilities of mMIMO depend on the quality of channel estimates. So far, we have assumed that $\tau_{\rm p} = K$ orthogonal pilot sequences are used for channel estimation. This is the common approach in mMIMO since it allows each BS to allocate orthogonal pilot sequences among its UEs, which are those originating the strongest interference. However, there might be use cases with stringent latency requirements in which only few samples $\tau_{\rm p}$ can be dedicated to channel estimation. In these cases, $\tau_{\rm p}$ will likely be smaller than $K$ and thus UEs within the same cell can be assigned to the same pilot sequence. This gives rise to intra-cell pilot contamination, which inevitably deteriorates the SE of mMIMO. We now investigate if NOMA can bring some benefits in these cases.

{\color{blue} Fig.~\ref{fig:SE_funcOftaup} depicts the sum UL SE as a function of number pilot signatures $\tau_{\rm p}$ with mMIMO and mMIMO-NOMA. We adopt the same setup of Fig.~\ref{SE_funcOfK_multicell}, where $K = 32$ UEs are equally distributed in four circle-areas of radius $r=20$\,m, and are randomly deployed in the cell area. Orthogonal spreading codes with length $N = 8$ are used for transmission and properly assigned to the different groups with mMIMO-NOMA thanks to Algorithm 2.} Fig.~\ref{fig:SE_funcOftaup} shows that SE starts reducing when $\tau_{\rm p} < 16$ with both mMIMO and mMIMO-NOMA. However, the decrease in performance is slightly lower with mMIMO-NOMA because it does not rely only on the quality of channel estimates for dealing with interference. Particularly, a large gain is observed with NOMA when MMSE is used with only one channel use (i.e., $\tau_{\rm p}=1$) for channel estimation. This is because MMSE is affected much from not having good channel estimates.

\begin{figure}[t!]
\centering\vspace{-0.7cm}
\begin{overpic}[unit=1mm,width=\linewidth]{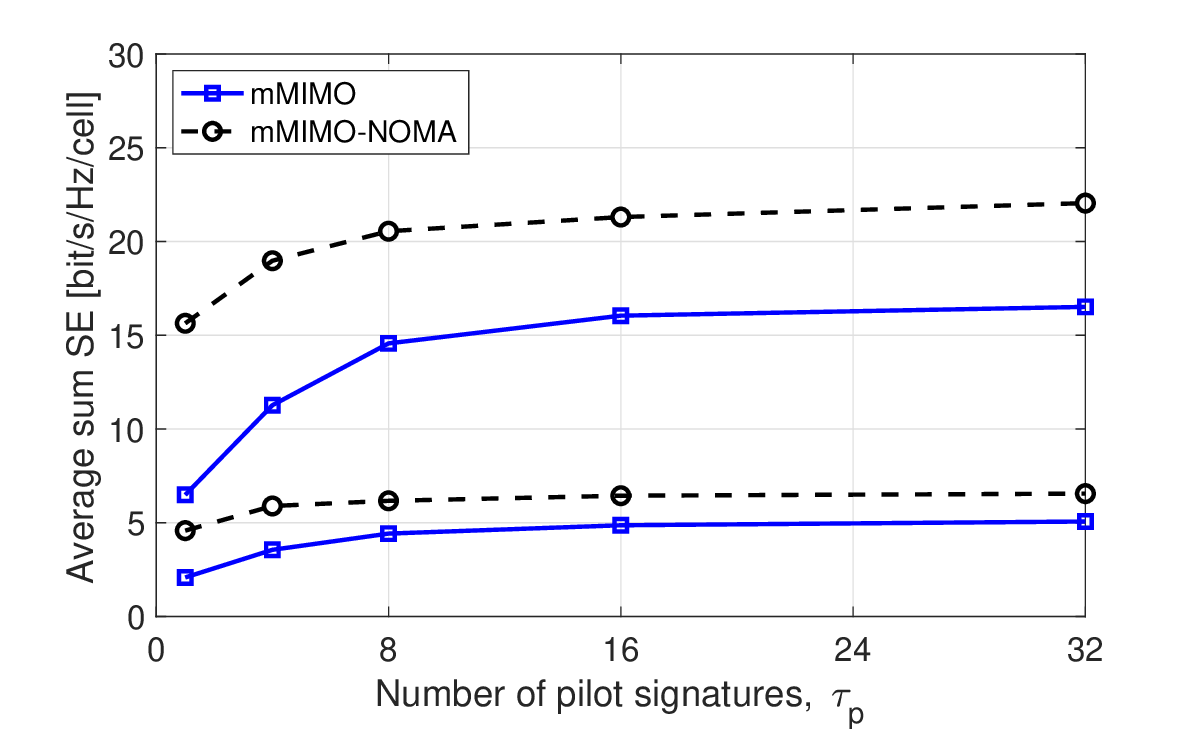}
		\put(55,39){\oval(5, 11)[r]}
	\put(62,50){\small MMSE}
	\put(63,49){\vector(-1, -1){5}}
		\put(26,26){\small MR}
	\put(32,26){\vector(1, -1){5}}
		\put(38,18){\oval(2, 6)[l]}
		\end{overpic}
\caption{{\color{blue} Average sum UL SE as a function of number pilot signatures $\tau_{\rm p}$ with mMIMO and mMIMO-NOMA.  $K = 32$ UEs and orthogonal spreading signatures of length $N = 8$ are considered.} }\vspace{-0.5cm}
\label{fig:SE_funcOftaup}
\end{figure}
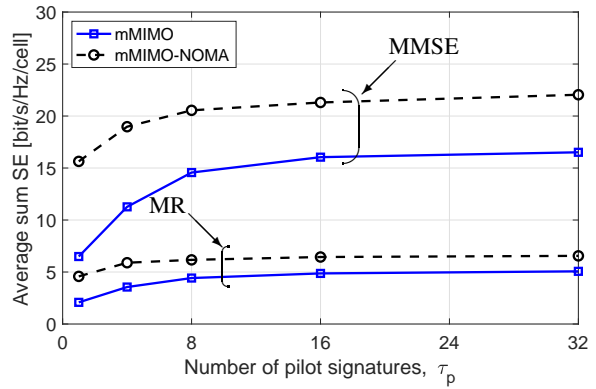


\section{Conclusions}\label{Sec:conclusion}
 {\color{blue} We investigated cases where code-domain NOMA can improve the spectral efficiency of mMIMO in the classical regime where $K < M$. Novel general SE expressions for arbitrary spreading signatures and combining/precoding schemes were provided. We used these expressions to show, by means of simulations, that the SE can be improved by NOMA in cases when poor favorable propagation conditions are experienced by the UEs. This may happen when the UEs are located close to each other and/or when planar arrays with insufficient resolution in the azimuth domain are considered.}
 
{\color{black}A two-step grouping algorithm was developed based on the k-means algorithm using the chordal distance as a similarity score metric to group the UEs with similar spatial correlation matrices. To fully take advantage of NOMA, the second step makes use of the Hungarian method to ensure that the $N-$length spreading sequences can be efficiently used for $N$ UEs per group}. {\color{blue} Numerical results showed that mMIMO-NOMA may provide some gains if spreading sequences are assigned to the UEs within the same group. This is valid, as expected, in the overloaded regime, but also even with the classical mMIMO setup, i.e. $M \gg K$. The analysis was carried out with orthogonal, random, and sparse spreading signatures, revealing that orthogonal spreading sequences are the best choice. We also showed that benefits can be achieved with NOMA when channel estimates of lower quality are available at the mMIMO BS. This can be of practical interest for massive machine type communications where short pilot sequences are generally used for channel estimation.}

\vspace{-0.1cm}
\appendices

\section{}
\label{app:proof:MMSE-estimate_h_jli}

The MMSE estimate of ${{\bf h}}_{li}^j$ is obtained as \cite{Kay:1993}
\begin{align}
\widehat{{\bf h}}_{li}^j = \mathbb{E}\left\{{{\bf h}}_{li}^j {\rm{vec}}\left({\bf Y}_{j}^p\right)^{\Htran}\right\}\left(\mathbb{E}\left\{{\rm{vec}}\left({\bf Y}_{j}^p\right) {\rm{vec}}\left({\bf Y}_{j}^p\right)^{\Htran}\right\}\right)^{-1} {\rm{vec}}\left({\bf Y}_{j}^p\right).\label{A.1}
\end{align}

By using ${\rm{vec}}\left({\bf ABC}\right) = \left({\bf C}^{T} \otimes {\bf A}\right){\rm{vec}}\left({\bf B}\right)$ we obtain
\begin{align}
\mathbb{E}\left\{{{\bf h}}_{li}^j {\rm{vec}}\left({\bf Y}_{j}^p\right)^{\Htran}\right\} & = \sqrt{p_{li}} \,{\vect{R}}_{li}^j\left({\bphiu}_{li}^{\Htran} \otimes {\bf I}_{M}\right) = \sqrt{p_{li}} \,\left({\bphiu}_{li}^{\Htran} \otimes {\vect{R}}_{li}^j\right)\label{A.2}
\end{align}
since the channels are independent. Similarly, one gets
\begin{align}\notag
&\mathbb{E}\left\{{\rm{vec}}\left({\bf Y}_{j}^p\right) {\rm{vec}}\left({\bf Y}_{j}^p\right)^{\Htran}\right\} \\
\notag & = \sum_{l^\prime=1}^{L}\sum_{i^\prime=1}^{K}p_{l^\prime i^\prime} \left( {\bphiu}_{l^\prime i^\prime} \otimes {\bf I}_{M}\right){\vect{R}}_{l^\prime i^\prime}^j \left( {\bphiu}_{l^\prime i^\prime}^{\Htran} \otimes {\bf I}_{M}\right) + \sigma^{2}{\bf I}_{M\tau_p} \\ \notag& = \sum_{l^\prime=1}^{L}\sum_{i^\prime=1}^{K} p_{l^\prime i^\prime} \left( {\bphiu}_{l^\prime i^\prime} \otimes {\bf I}_{M}\right)\left({\bphiu}_{l^\prime i^\prime}^{\Htran} \otimes {\vect{R}}_{l^\prime i^\prime}^j\right) + \sigma^{2}{\bf I}_{M\tau_p} \\ & = \sum_{l^\prime=1}^{L}\sum_{i^\prime=1}^{K} p_{l^\prime i^\prime}  \left({\bphiu}_{l^\prime i^\prime}{\bphiu}_{l^\prime i^\prime}^{\Htran} \right)\otimes {\vect{R}}_{l^\prime i^\prime}^j+ \sigma^{2}{\bf I}_{M\tau_p}.\label{A.3.4}
\end{align}
By substituting \eqref{A.2} and \eqref{A.3.4} into \eqref{A.1} yields \eqref{eq:MMSEestimator_h_jli}. \vspace{-0.2cm}

\section*{Acknowledgment}
The authors would like to acknowledge Jakob Hoydis for useful discussions in the development of Algorithm 1.\vspace{-0.2cm}

\bibliographystyle{IEEEtran}
\bibliography{IEEEabrv,refs,ref,ref_book}

\end{document}